\documentclass{article}

\usepackage{microtype}
\usepackage{graphicx}
\usepackage{subfigure}
\usepackage{booktabs} 

\usepackage{hyperref}

\usepackage[arxiv]{icml2025}

\usepackage{amsmath}
\usepackage{amssymb}
\usepackage{mathtools}
\usepackage{amsthm}

\usepackage[capitalize,noabbrev]{cleveref}

\usepackage{threeparttable}

\theoremstyle{plain}

\theoremstyle{definition}

\theoremstyle{remark}

\usepackage[textsize=tiny]{todonotes}

\begin{document}

\twocolumn[
\icmltitle{In Pursuit of Predictive Models of Human Preferences Toward AI Teammates}



\icmlsetsymbol{equal}{*}

\begin{icmlauthorlist}
\icmlauthor{Ho Chit Siu}{mitll}
\icmlauthor{Jaime D. Pe\~na}{mitll}
\icmlauthor{Yutai Zhou}{mitll,usc}
\icmlauthor{Ross E. Allen}{mitll}
\end{icmlauthorlist}

\icmlaffiliation{mitll}{Lincoln Laboratory, Massachusetts Institute of Technology, Lexington, MA, USA}
\icmlaffiliation{usc}{University of Southern California, Los Angeles, CA, USA}

\icmlcorrespondingauthor{Ross E. Allen}{ross.allen@ll.mit.edu}

\icmlkeywords{Machine Learning, ICML}

\vskip 0.3in
]



\printAffiliationsAndNotice{}  

\begin{abstract}
We seek measurable properties of AI agents that make them better or worse teammates from the subjective perspective of human collaborators.
Our experiments use the cooperative card game Hanabi---a common benchmark for AI-teaming research.
We first evaluate AI agents on a set of objective metrics based on task performance, information theory, and game theory, which are measurable without human interaction. 
Next, we evaluate subjective human preferences toward AI teammates in a large-scale (N=241) human-AI teaming experiment.
Finally, we correlate the AI-only objective metrics with the human subjective preferences. 
Our results refute common assumptions from prior literature on reinforcement learning, revealing new correlations between AI behaviors and human preferences. %
We find that the final game score a human-AI team achieves is less predictive of human preferences than esoteric measures of AI action diversity, strategic dominance, and ability to team with other AI.
In the future, these correlations may help shape reward functions for training human-collaborative AI.
\end{abstract}

\section{Introduction}\label{sec:introduction}

Reinforcement learning (RL)~\cite{sutton2018reinforcement} has been the foundation of almost every artificial intelligence (AI) that has achieved superhuman capabilities in \emph{competitive} tasks such as computer games \cite{mnih2013playing,berner2019dota,vinyals2019grandmaster,wurman2022outracing}, board games \cite{silver2018general}, or drone racing \cite{kaufmann2023champion}. %
The recent proliferation of systems such as large language models and autonomous cars marks a new chapter in the history of AI that features \emph{human-collaborative} technologies designed to help humans with tasks from the mundane (e.g. drafting emails) to the dramatic (e.g. avoiding life-threatening collisions). %
Although these collaborative AIs contrast with their competitive ancestors, we see that reinforcement learning continues to play an essential role in their development~\cite{rafailov2024direct,ziegler2019fine,kiran2021deep}. %
Fundamental differences exist in the reward structures needed to train collaborative AI; that is, in order for the AI to be a ``good'' collaborator, training rewards must align with the subjective preferences of their human teammates. Unfortunately, using traditional RL techniques that require vast amounts of training time, it is almost never practical to have humans provide direct feedback on AI behavior during the training process~\cite{christiano2017deep}. 
Therefore, in this work, we seek objective measures of AI behavior that are predictive of subjective human preferences toward AI collaborators. In the long term, finding such objective metrics could then serve as a human-proxy reward function during RL training.

\emph{But}, some readers may ask, 
\emph{do people not simply want a teammate that helps them do better at the task at hand; that is to score more goals \cite{kurach2020google}, prepare more meals \cite{carroll2019utility}, or minimize medical misdiagnoses \cite{suzuki2017overview,richens2020improving}?} %
Our results, however, indicate that maximizing objective task performance may only be a weak predictor of a human's preferences toward AI teammates, and that other objective measures based on information theory and game theory may be more predictive of how humans will perceive AI teammates. 

These conclusions may not be surprising from a behavioral economics perspective where human preference is often shown to be nuanced and counterintuitive~\cite{thaler2009nudge}. We argue, however, that quantification and analysis of such preferences in the domain of human-AI collaboration is understudied and we urge note to be taken by the reinforcement learning community, where much of the literature originates from competitive domains in which reward functions emphasize simply ``out scoring'' an opponent and superhuman capabilities can be learned via \emph{self-play} without human interaction~\cite{silver2018general}. 

Existing works have arrived at similar conclusions regarding the primacy---or lack thereof---of task performance as the driver for human preferences. \citet{tauer1999winning} provides a human-subjects study of how enjoyment and intrinsic motivation is impacted by competition and achievement orientation; however, it reveals little about human preferences toward collaborative teammates or how to align AI behaviors with those preferences. %
\citet{zhao2020winning} explores how RL can be used in video game design to create AI players for play testing and evaluation---including a case study on training an assistive AI that complements teammate's play styles---rather than just agents that ``beat the game''. However, the case study lacks any human-subjects experiments that actually evaluate human preferences toward such an assistive AI.

Other literature has made significant contributions in using RL to align AI behaviors to human preferences, particularly large language models~\cite{christiano2017deep,ziegler2019fine,rafailov2024direct}. However, these works largely focus on the alignment algorithms given a dataset of preferences regarding a particular language model; they have less to say on the identification or prediction of such preferences in the first place. 

We hypothesize that there exist objective metrics of AI behavior that are: 1) measurable without human interaction, and 2) predictive of human preferences toward such AI when introduced as a collaborator. %

In our experiments, we first collect a set of existing AI agents developed for a particular task; in this case, a collaborative card game where many of the agents have been trained using reinforcement learning (RL). Next, we conduct AI-only teaming experiments whereby pairs of AI agents collaborate on the task over many trials. With this AI-only dataset we make objective measurements of the AI's behaviors with metrics based on task performance, information theory, and game theory. We then downselect a subset of our AI pool for experimentation with human teammates. During our human-AI teaming experiments, human participants are asked to provide feedback about their AI teammates in terms of preference, trust, predictability, etc. Finally we attempt to correlate an AI's objective metrics from AI-only experiments with its subjective metrics from human-AI experiments. This correlation analysis is the primary contribution of our work as it may inform future works in human-AI teaming how to better shape reward functions while training agents without humans-in-the-loop.

We use the collaborative card game \emph{Hanabi}, a key benchmark in the research of human-AI teaming~\cite{bard2020hanabi,canaan2020generating,hu2020other,siu2021evaluation}. %
In Hanabi, players work together to stack a set of cards in appropriate order, similar to Solitaire, but with multiple players and limited information sharing (detailed rules in the Appendix~\ref{app:hanabi_rules}). %
What sets Hanabi apart from many collaborative games~\cite{carroll2019utility,strouse2021collaborating,gray2020human,meta2022human,cooke2004synthetic} is players' inability to see their own cards and the strictly limited communication between players. %
Players cannot communicate other than using a small set of ``hints'' provided to the team during their turn (e.g. pointing out all cards of a certain color).
The highly-restricted nature of Hanabi---which eliminates many confounding variables of human bias when interacting with AI~\cite{rheu2021systematic,hoff2015trust,oneill2022human,guznov2020robot}---helps us search for innate/algorithmic forms of coordination that humans find desirable in the absence of other communication factors. 

We see several counter-intuitive results: we find that the final score the human-AI team achieves is only a weak predictor of the human participant's affinity toward the AI (Section~\ref{subsec:sub_obj_metric_correlation}). This is particularly noteworthy in Hanabi where limited communication leaves little other than the team's performance upon which a person might form an opinion of teammates. This result bolsters a similar observation previously made by Siu et al. \cite{siu2021evaluation}. 

Further results cast doubt on the utility of some of the most widely-used techniques in reinforcement learning (RL) when applied to human-AI teaming problems. Techniques like \emph{self-play}~\cite{bansal2017emergent} and \emph{cross-play}~\cite{hu2020other,lucas2022any,lupu2021trajectory,strouse2021collaborating,nekoei2021continuous}---where AI agents are trained by playing with copies and near-copies of themselves---have been foundational to almost every major RL milestone in recent years~\cite{silver2018general,berner2019dota,vinyals2019grandmaster,wurman2022outracing}. However, our results show that an AI's self-play and cross-play scores may only be weak-to-moderate predictors of human preference toward the AI (Section~\ref{subsec:sub_obj_metric_correlation}). %


We see evidence that AI training should not only maximize the task performance, but also avoid taking actions that may be \textit{perceived} as irrational/sub-optimal by a human (Section~\ref{subsec:sub_obj_metric_correlation}). %

In the following sections we present the key methods and results from each phase of our experiments, including: selection of objective metrics for AI behavior (Section~\ref{subsec:results_objective_metrics_selection}); selection of AI Hanabi agents for which objective metrics are evaluated in AI-only experiments and how this informs our human-AI experimental procedure (Section~\ref{subsec:ai-only-evaluations}); selection and evaluation of subjective metrics in human-AI experiments (Section~\ref{subsec:human_experiments_methods}); and finally, correlations between objective AI-only metrics and subjective human-AI metrics (Sections~\ref{subsec:sub_obj_metric_correlation}). In Section~\ref{sec:discussion} we interpret our results and discuss limitations and future work.


\section{AI-only Experiments}
\label{sec:ai_only_experiments}

\subsection{Objective Metrics Selection}
\label{subsec:results_objective_metrics_selection}


We seek objective metrics of AI behavior that predict subjective preferences of humans, which are impractical to directly feedback as reward functions during the RL training processes~\cite{christiano2017deep}. %
Existing literature~\cite{siu2021evaluation} provided evidence that humans may not prefer an AI just because they achieve a higher score (i.e. task performance) as a team; therefore, we also consider information-theoretic and game-theoretic objective measures of AI behavior (Table~\ref{tab:agent_metrics}). Metrics we considered, but excluded, are described in Appendix~\ref{subapp:excluded_metrics}.

\begin{table*}[ht]
\footnotesize
\caption{Objective evaluation metrics for Hanabi agents.}
\label{tab:agent_metrics}
\centering
\begin{tabular}{l|l|l}
\hline
Name & Type& Notes \\
\hline
Human-AI Game Score & Task Performance & requires human-in-the-loop \\
Self-Play Score \cite{lucas2022any} & Task Performance& \\
Intra-XP Score \cite{lucas2022any} & Task Performance& only usable if different versions of the agent exist\\
Inter-XP Score \cite{lucas2022any} & Task Performance& depends on the agents in the experiment pool\\
Action Distribution (AD) Entropy & Information Theory& \\
Action-Response Distribution (ARD) Entropy & Information Theory& \\
Instantaneous Coordination (IC) \cite{jaques2019social} & Information Theory&  \\
Context Independence (CI) \cite{bogin2018emergence} & Information Theory& domain-specific adaptations \\ 
Dominated Action 1 (G1) Frequency & Game Theory& frequency of discarding a known-to-be playable card \\
Dominated Action 2 (G2) Frequency & Game Theory& frequency of playing a known-to-be unplayable card \\
Dominant Action (G3) Frequency & Game Theory& frequency of playing a known-to-be playable card \\

\end{tabular}
\end{table*}

\emph{Task Performance} metrics measure performance on task-specific quantities; i.e. \emph{payoffs} like the number of goals scored in a soccer game. Hanabi performance is the final game score achieved by the team which is an integer between 0 and 25~\cite{bard2020hanabi}. %
The human-AI game score requires human interaction, so while it is an objective metric, it does not scale well to AI training, and is provided as a reference for other metrics.
The mathematical definitions of self-play, intra-algorithm cross-play (intra-XP), and inter-algorithm cross-play (inter-XP) scoring metrics are given by \citet{lucas2022any} and restated in the Appendix~\ref{subapp:task_performance_objective_metrics_selection}. In short, they measure the average score of an agent when paired with: a copy of itself (self-play), a distinct agent trained with the same algorithm (intra-XP), and a pool of agents trained/developed with arbitrary/unrelated algorithms (inter-XP)~\cite{hu2020other,lucas2022any,nekoei2021continuous}. %
These metrics allow us to test a hypothesis implied in existing literature~\cite{silver2018general,hu2020other,lucas2022any,lupu2021trajectory} that task performance predicts human preferences toward AI teammates.

\emph{Information Theory}-based metrics consider distributional properties of an agent's actions such as entropy and mutual information~\citep[Chp~2]{cover2006elements}. These metrics, rigorously defined in the Appendix~\ref{subapp:information_theoretic_metric_defs}, arise from literature on emergent communication~\cite{lowe2019pitfalls}, and allow us to test the hypothesis that the patterns of an agent's actions---and perhaps some forms of latent communication from such patterns---influence a human's preferences.

\emph{Game theory}-based metrics in Table~\ref{tab:agent_metrics} include the frequency of an agent executing dominated (G1 and G2) and dominant (G3) actions. %
As a combination of the concepts of \emph{bounded rationality}---which describes human decision-making processes that are limited by our information processing capability ~\cite{simon1990bounded}---and \emph{strategic dominance}~\citep[Chp~4]{tadelis2013game}, these metrics measure how often an agent takes actions that are short-term optimal (G3) 
or irrational (G1 and G2). 
A \textit{dominated} strategy is one that results in a worse outcome for a player, regardless of what other players do, and \textit{dominant} strategy is one that results in a better outcome regardless of what other players do. We restrict our investigation to single-move contexts, and thus refer to dominated and dominant \textit{moves}. %
See the Appendix~\ref{subapp:game_theoretic_metric_defs} for strict definitions. %
With game-theoretic metrics, we can test the hypothesis that humans prefer agents that act in a fashion that is perceived to be rational and disfavor those that are perceived to act irrationally. 

\subsection{Objective Metrics Evaluation}
\label{subsec:ai-only-evaluations}

To evaluate the objective metrics (Sec~\ref{subsec:results_objective_metrics_selection}), we conducted a series of AI-only experiments, consisting of  1) collecting a broad range of publicly-available Hanabi AI agents; 2) having all agent pairs play games together; 3) evaluating the metrics (Table~\ref{tab:agent_metrics}) for each agent; and 4) down-selecting agents for human-AI experiments. Each distinct agent played 125 two-player games with every other agent identified in Table~\ref{tab:bots_evaluated} (i.e. 125 games per agent pairing permutations). All AI-only experiments were carried out on a single Intel Xeon E5-2683 compute node (28 CPU cores, 2.0Ghz) with 256GB of RAM.

All of the AI agents in Table~\ref{tab:bots_evaluated} were evaluated for objective metrics in Tables~\ref{tab:agent_metrics}; however, due to time constraints from the number of games needed for statistical power, not all agents could be tested as teammates within human experiments (Section~\ref{sec:human_ai_experiments}). We down-selected the AI pool for human experimentation by attempting to span the objective metric space; agents selected for human experimentation are indicated with a star in Table~\ref{tab:bots_evaluated} and this subset of objective metrics are summarized in Table~\ref{tab:ai_only_summary}. 

We categorized Hanabi AI agents based on the source of their behavior: handcrafted rules, search, and reinforcement learning (Table \ref{tab:bots_evaluated}). Agents that imitate humans (e.g. CloneBot~\cite{lerer2020improving} and piKL3~\cite{hu2022human}) were excluded because the models were not made publicly available by the respective authors. %

\begin{table*}[ht]
\footnotesize
\centering
\begin{threeparttable}
\caption{Hanabi AI agents selected for objective metric evaluation.
}
\label{tab:bots_evaluated}
\centering
\begin{tabular}{l|l|l}
\hline
Name & Type & Description \\
\hline
RandomBot* & Rule & Randomly chooses any legal move \\
SimpleBot \cite{odwyer2019hanabi} & Rule & Plays playable cards and hints about playable cards \\
ValueBot \cite{odwyer2019hanabi} & Rule & SimpleBot rules + hints about valuable cards \\
HolmesBot* \cite{odwyer2019hanabi} & Rule & ValueBot rules + inference on own hand, uses bombs \\
SmartBot* \cite{odwyer2019hanabi} & Rule & Uses common human conventions, prefers to play cards its partner \\
& & doesn't know it knows \\
SPARTA \cite{lerer2020improving} & Search & Monte Carlo search over previously agreed-upon convention \\
IQL \cite{hu2019simplified} & RL & Recurrent deep Q learning \\
VDN \cite{hu2019simplified} & RL & IQL agent that also learns a joint Q function \\
SAD(+Aux) \cite{hu2019simplified} & RL & State-of-the-art in self-play-trained agents \\ 
Div(+Aux) \cite{hu2020other} & RL & SAD(+Aux) using a diverse set of network architectures (originally\\
& & used as a comparison against Other-Play) \\
Other-Play(+Aux*) \cite{hu2020other} & RL (+Search) & Designed to be robust to lack of coordination on how to break symmetries\\
& & in communication\\
Any-Play(+Aux*) \cite{lucas2022any} & RL (+Search) & Diversity-based intrinsic auxiliary reward\\
Off-Belief* \cite{hu2021off} & RL & Based on (potentially iterated applications of) an approximately optimal\\
& & grounded policy without conventions\\
\end{tabular}
\begin{tablenotes}
    \item (+Aux) indicates the use of both the base agent and a version trained with the auxiliary task of predicting whether a card is playable, discardable, or unknown \cite{hu2019simplified}. Off-Belief (OBL) included agents with one to four iterations of optimal policies, where each iteration assumes one's partner plays with the previous iteration's policy (OBL1 to OBL4) \cite{hu2021off}. Starred agents, including OBL levels 1, 2, and 4 were used in human-AI experiments.
\end{tablenotes}
\end{threeparttable}
\end{table*}


RandomBot, SmartBot, and HolmesBot are all rule-based agents, while the others use neural networks trained with various reinforcement learning algorithms; the +Aux agents also use a search-based heuristic with the RL policy~\cite{hu2019simplified,hu2020other,lucas2022any}. %
RandomBot is included as a boundary case for many of the objective metrics.



SmartBot was the highest-performing of the rule-based bot pool, while HolmesBot represented the culmination of the line of work starting from SimpleBot through ValueBot; see \citet{siu2021evaluation} for in-depth comparison of rule-based agents. AP+Aux was included due to its high inter-XP performance ~\cite{lucas2022any}; additionally its  training is similar to fictitious co-play, a method found to be effective for working with humans in the game Overcooked \cite{strouse2021collaborating}. Other-Play+Aux was included as a reference point for a bot known to have high AI-only performance, but poor human ratings \cite{hu2020other,siu2021evaluation}.

OBL1 approximates purely grounded play---behavior based solely on game information, and not on other agents' behavior. 
OBL2 is analogous to first-level theory of mind, i.e. it assumes that its teammate plays in a grounded manner. OBL4 was also included as the highest-performing OBL agent across multiple objective metrics (see Table~\ref{tab:ai_only_summary}) \cite{hu2021off}.


\begin{table*}[ht]
\scriptsize
\centering
\caption{Objective metrics in AI-only evaluations for agents used in human-AI experiments.}
\label{tab:ai_only_summary}
\centering

\begin{threeparttable}
\begin{tabular}{l|rrrrr}
\hline
Agent & Self-play & Intra-XP & Inter-XP & CI & IC \\
\hline
AP+Aux & $23.128 \pm 1.743$ & $22.000 \pm 2.385$ & $16.125 \pm 6.805$ & $0.107 \pm 0.005$ & $0.442 \pm 0.063$ \\
HolmesBot & $20.684 \pm 2.667$ &  - & $6.197 \pm 4.056$ & $0.118 \pm 0.007$ & $0.450 \pm 0.029$ \\
OBL1 & $21.268 \pm 2.001$ & $21.168 \pm 2.297$ & $12.535 \pm 6.984$ & $0.103 \pm 0.005$ & $0.163 \pm 0.023$\\
OBL2 & $23.428 \pm 1.606$ & $23.301 \pm 1.655$ & $14.822 \pm 7.759$ & $0.104 \pm 0.006$ & $0.271 \pm 0.027$\\
OBL4 & $\mathbf{24.168 \pm 1.215}$ & $\mathbf{23.927 \pm 1.550}$ & $13.539 \pm 8.216$ & $0.103 \pm 0.004$ & $0.329 \pm 0.033$\\
OP+Aux & $23.896 \pm 1.687$ & $23.002 \pm 2.532$ & $\mathbf{16.402 \pm 6.949}$ & $0.105 \pm 0.007$ & $0.357 \pm 0.040$\\
RandomBot & $1.180 \pm 1.231$ &  - & $1.641 \pm 1.561$ & $0.101 \pm 0.005$ & $0.023 \pm 0.011$\\
SmartBot & $23.288 \pm 1.795$ &  - & $10.393 \pm 6.409$ & $0.105 \pm 0.006$ & $0.370 \pm 0.047$\\
\end{tabular}
\begin{tabular}{l|rrrrr}
\hline
\hline
Agent & AD-entropy & ARD-entropy & G1-dominated & G2-dominated & G3-dominant\\
\hline
AP+Aux & $2.653 \pm 0.099$ & $4.579 \pm 0.262$ & $0.002 \pm 0.002$ & $0.012 \pm 0.006$ & $0.097 \pm 0.031 $\\
HolmesBot & $2.560 \pm 0.061$ & $4.271 \pm 0.169$ & $0.005 \pm 0.004$ & $\mathbf{0.002 \pm 0.003}$ & $0.067 \pm 0.027 $\\
OBL1 & $2.760 \pm 0.104$ & $4.763 \pm 0.376$ & $0.001 \pm 0.001$ & $0.008 \pm 0.007$ & $\mathbf{0.131 \pm 0.070} $\\
OBL2 & $2.821 \pm 0.050$ & $4.736 \pm 0.369$ & $0.001 \pm 0.001$ & $0.010 \pm 0.007$ & $0.119 \pm 0.057 $\\
OBL4 & $2.800 \pm 0.041$ & $4.727 \pm 0.356$ & $\mathbf{0.000 \pm 0.001}$ & $0.010 \pm 0.008$ & $0.106 \pm 0.051 $\\
OP+Aux & $2.737 \pm 0.073$ & $4.688 \pm 0.300$ & $0.001 \pm 0.001$ & $0.012 \pm 0.006$ & $0.109 \pm 0.047 $\\
RandomBot & $2.912 \pm 0.011$ & $4.109 \pm 0.315$ & $0.041 \pm 0.010$ & $\mathbf{0.001 \pm 0.001}$ & $0.045 \pm 0.009 $\\
SmartBot & $2.727 \pm 0.060$ & $4.636 \pm 0.281$ & $0.004 \pm 0.002$ & $0.008 \pm 0.006$ & $0.092 \pm 0.034 $\\
\end{tabular}
\begin{tablenotes}
    \item Results in bold indicate best-performing agent per the respective metric. For information-theoretic metrics of CI, IC, AD-entropy, and ARD-entropy, no agents are highlighted because it is not obvious if higher or lower values should be labeled as ``better'' performance; analysis and discussion of such performance is left to Section~\ref{subsec:sub_obj_metric_correlation}.
\end{tablenotes}
\end{threeparttable}

\end{table*}


Several results are worth noting in Table~\ref{tab:ai_only_summary}. OBL4 agents are the highest performing in terms of task performance metrics of self-play and intra-XP scores, while OP+Aux and AP+Aux are highest performing in inter-XP; this largely reproduces numerical results presented in previous literature~\cite{hu2021off,lucas2022any}. %
OBL4 produces the lowest frequency of dominated actions (G1) while OBL1 produces the highest frequency of dominant actions (G3). RandomBot and HolmesBot produce the lowest ARD-entropy and lowest frequency of G2-dominated actions, but this is likely due to the bots losing games very early, leading to a reduced range of game states experienced (reducing entropy) and few situations where they have the opportunity to play a card that is known to be unplayable (reducing G2-dominated frequency).

\section{Human-AI Experiments}
\label{sec:human_ai_experiments}

\subsection{Subjective Metric Selection \& Evaluation}
\label{subsec:human_experiments_methods}

To evaluate human subjective preferences, we conducted human-AI teaming experiments using an instance of each agent type listed in Table~\ref{tab:ai_only_summary} and a pool of human participants ($N=241$ participants). %

After their consent process, participants were introduced to the web-based experiment, hosted on an AWS EC2 instance, and shown a video and static content outlining the rules of Hanabi along with the experiment graphical user interface. Next, participants answered a set of questions about the game state of provided GUI snapshots to ensure comprehension. Answers and explanations were provided after each question, and participants were screened out if too many answers were incorrect. Participants then answered a question rating their own level of familiarity with Hanabi before the experiment, and the nature of their previous experience.

Each participant played eight two-player games, four with one Hanabi AI, followed by four with another. Only the last three games with each agent were used for analysis. The selection and order of AI agents was randomized for each participant. Participants were blind as to which AI agents they were playing with, referring to them simply as the ``first bot'' or ``second bot'' with which they interacted.


\begin{table*}[ht]
\footnotesize
\centering
  \begin{threeparttable}      
    \centering
    \caption{Likert item statements after a block of games with the same bot (left) and after all games with both bots (right).}
    \label{tab:post-bot-left-experiment-right}
    \begin{tabular}{ l | l || l | l }
      \toprule
      \multicolumn{2}{c}{\bfseries Post-block Likert Statements}           & \multicolumn{2}{c}{\bfseries Post-experiment Likert Questions} \\
      \midrule
      B1& I am playing well \hspace{5cm}                    & P1     & Which bot did you prefer playing with?       \\
      B2& The bot is playing well               & P2     & Which bot did you trust more?                \\
      B3& The bot and I have good teamwork      & P3     & Which bot did you understand more?           \\
      B4& I understand the bot’s intentions     & P4     & Which bot understood you more?               \\
      B5& The bot understands my intentions     & P5     & Which bot was the better Hanabi player?      \\
      B6& I trust the bot to do the right thing & P6     & Which bot was more predictable?              \\
      B7& The bot is predictable                & P7     & Which bot was easier to work with?           \\
      B8&  \vtop{\hbox{\strut The team would perform well}\hbox{\strut in a future game }}  &   &       \\         
      \bottomrule
    \end{tabular}
    \begin{tablenotes}
        \item Post-block statements followed the question “Based on your “actual” games, how would you rate your level of agreement with the following statements for games with this bot?”
    Responses were in the form of a 7-point Likert scale ranging from Strongly Disagree to Strongly Agree. Post-experiment questions had responses in the form of a 7-point Likert scale ranging from “Definitely the first bot” to “Definitely the second bot”.
    \end{tablenotes}
    \end{threeparttable}

\end{table*}

After each block of four games, participants rated that block's AI teammate on eight 7-point Likert scale questions regarding performance and teamwork (Table \ref{tab:post-bot-left-experiment-right}, left), which were largely derived from \citet{hoffman2019evaluating}. After both blocks of games, participants directly compared their two AI teammates on seven 7-point Likert scale questions (Table \ref{tab:post-bot-left-experiment-right}, right). Answers for questions B3 to B8 in Table \ref{tab:post-bot-left-experiment-right} (left) were then summed to compute an overall \textit{teamwork rating}---producing an integer value between 0-35---which is the basis for much of our subjective metric analysis. %

To probe which moves potentially had the greatest negative effects on perception, we added a button labeled ``Partner's Last Move was Questionable'' to the interface. Participants were instructed to click the button whenever they perceived their partner's last move as ``questionable,'' with the understanding that clicking it would not change their partner's behavior. We intentionally did not define or provide examples for such moves to avoid biasing participants.


Participants were recruited via MIT Behavioral Research Lab's (BRL) subject pool, and emails to university mailing lists. 245 adult participants completed the experiment, though server issues prevented the game scores of four of them from being recorded, so we used the remainder (241) in our analysis. 
The protocol was approved by the MIT Committee on the Use of Humans as Experimental Subjects (COUHES) as a minimal-risk experiment under Exempt Research. All subjects provided written, informed consent and all experiments were carried out in accordance with the relevant guidelines and regulations. No significant risks were expected or reported and we did not collect any personally-identifiable information.


Participants received a \$5 USD gift card for experiment completion and an additional \$1 for each point they scored on average across the non-``test'' games of both bots (maximum of \$30 total); subjects were told that their ``test'' games did not count towards payment. Experiments took approximately 30-90 minutes to complete, with experiment length generally scaling with performance. 



\begin{figure*}[ht]
  \centering
  \includegraphics[width=0.7\linewidth]{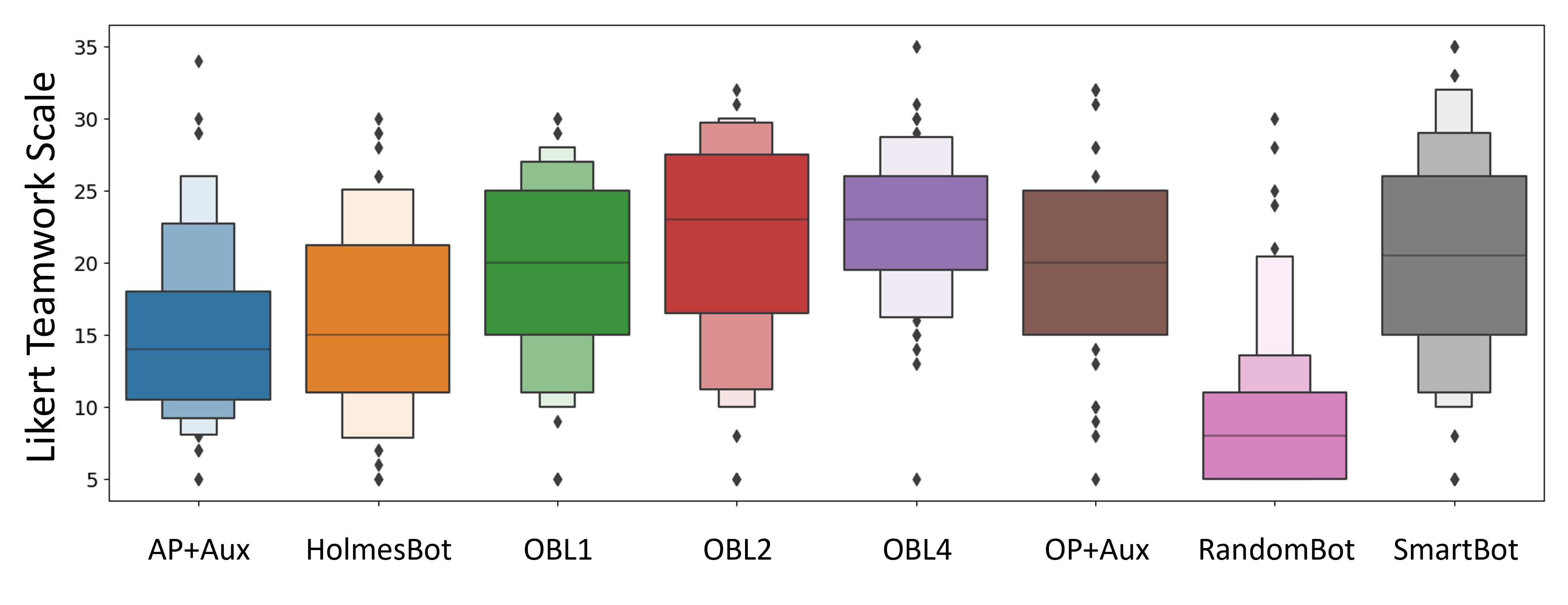}
  \caption{Letter-value plot of subjective teamwork rating statistics for each AI agent used in human-AI teaming experiments.}
  \label{fig:hai_teamwork_ratings}
\end{figure*}

Figure~\ref{fig:hai_teamwork_ratings} shows teamwork rating statistics for each agent type. As expected, RandomBot is the lowest rated agent.
SmartBot and OP+Aux have comparable teamwork ratings which largely supports past literature that found the same relationship~\cite{siu2021evaluation}. %
Of particular interest is the low rating of OP+Aux and AP+Aux agents. This tends to refute the assumption made by \citet{lucas2022any} that agents with higher inter-XP scores will naturally be preferred by human teammates. %
Indeed, results in Section~\ref{subsec:sub_obj_metric_correlation} will indicate that inter-XP does not have a statistically significant correlation to teamwork rating. 

OBL-based agents achieved some of the highest teamwork ratings from human participants, which is noteworthy for two reasons. %
First, it helps settle an open question from \citet{siu2021evaluation} by showing that learning-based AI teammates can be developed that are preferred over rule-based agents. %
Second, the same agents achieving high teamwork ratings also achieved some of the highest objective metrics in Table~\ref{tab:ai_only_summary}. However, this does not prove a correlation between the two; such analysis is left to Section~\ref{subsec:sub_obj_metric_correlation}.




\subsection{Objective-Subjective Metric Correlations}
\label{subsec:sub_obj_metric_correlation}

We now correlate the objective metrics of AI behavior (Sections~\ref{subsec:ai-only-evaluations}, independent variable) with the human-assessed subjective teamwork rating (Section~\ref{subsec:human_experiments_methods}, dependent variable).  Table~\ref{tab:regressions} summarizes these correlation results. %
We report correlations both with and without RandomBot because RandomBot is often an outlier for objective metrics.

In Table~\ref{tab:regressions}, $p$ indicates the significance of the regression, $r$ is the strength and direction of the correlation, and $m$ is the slope of the linear correlation. $m$ must be interpreted in the context of each variable, and can be thought of as ``for each unit of change in objective AI metric, how much does the subjective teamwork rating change?''

\begin{table*}[h]
\footnotesize
\centering
\begin{threeparttable}
\caption{Summary of correlation analysis between objective metrics of AI behavior (independent variable) and subjective teamwork rating by humans toward the AI (dependent variable). See Section \ref{subsec:sub_obj_metric_correlation} for further details.}
\label{tab:regressions}
\centering
\begin{tabular}{l|rrr|rrr}
\hline
& \multicolumn{6}{c}{\bf{Correlation w/ Subjective Teamwork Rating}} \\
\bf{Objective AI Behavior Metric \hspace{4cm}} & & \\
& \multicolumn{3}{c}{including RandomBot} & \multicolumn{3}{c}{excluding RandomBot} \\
&   r & m & p & r & m & p \\
\hline
Human-AI Game Score Mean &  0.350 & 0.365 & 0.000 & 0.203 & 0.204 & 0.000 \\
Self-Play Score Mean &  \bf{0.447} &  \bf{0.461} & \bf{0.000} &        0.168 &        1.049 &        0.001 \\
Intra-XP Score Mean* &  - &    - & - &        \bf{0.234} &        \bf{1.648} &        \bf{0.000} \\
Inter-XP Score Mean &  0.370 &    0.600 & 0.000 &        \textcolor{gray}{0.047} &        \textcolor{gray}{0.109} &        \textcolor{gray}{0.344} \\
AD-Entropy Mean & \textcolor{gray}{-0.086} &   \textcolor{gray}{-6.805} & \textcolor{gray}{0.060} &        \bf{0.301} &       \bf{27.199} &        \bf{0.000} \\
ARD-Entropy Mean &  \bf{0.466} &   \bf{15.899} & \bf{0.000} &        0.233 &       11.505 &        0.000 \\
CI Mean & \textcolor{gray}{-0.028} &  \textcolor{gray}{-47.543} & \textcolor{gray}{0.542} &       -0.218 &     -340.670 &        0.000 \\
IC Mean** &  \bf{0.517} &      - &   - &        \bf{0.332} &          - &          - \\
G1-dominated: Discard Playable & \bf{-0.447} & \bf{-253.569} & \bf{0.000} &       -0.200 &     -902.899 &        0.000 \\
G2-dominated: Play Unplayable &  0.332 &  666.770 & 0.000 &        \textcolor{gray}{0.058} &      \textcolor{gray}{137.098} &        \textcolor{gray}{0.242} \\
G3-dominant: Play Playable &  0.421 &  121.216 & 0.000 &        0.166 &       63.551 &        0.001 \\
\end{tabular}
\begin{tablenotes}
\item r is the correlation coefficient, and m is the slope of the regression line relating objective AI behavior metric (independent) and subjective teamwork rating (dependent), and the significance threshold is $p < \alpha=0.05/18=0.00278$. Correlation values are presented with both inclusion and exclusion of RandomBot from the dataset due to it being an outlier in many objective metrics (see Table~\ref{tab:ai_only_summary}). Objective metrics with the strongest correlations to teamwork rating are bold face, and metrics with statistically insignificant correlations are grayed-out. *Intra-algorithm cross-play score excludes all non-trainable, rule-based bot types (RandomBot, HolmesBot, and SmartBot). **IC used a nonlinear fit and does not have m and p values.
\end{tablenotes}
\end{threeparttable}
\end{table*}

Perhaps the most surprising result is that human-AI game score is a weaker predictor of the subjective teamwork rating (lower correlation value) than many of the other objective metrics, such as self-play (including RandomBot) and intra-XP (excluding Randombot). 
In other words, when trying to predict how a human will rate an AI as a teammate, it is better to know the average score an AI achieves while teaming with \emph{other AI} than it is to know the average score when teaming with \emph{actual humans}! %

Task performance metrics (self-play, intra-XP, inter-XP) \emph{do not} exhibit the strongest correlations with subjective teamwork rating. Indeed, information-theoretic metrics such as action distribution entropy (AD-entropy), action-response distribution entropy (ARD-entropy) and instantaneous coordination (IC) are more strongly correlated with the human-provided teamwork rating. Game-theoretic metrics, such as frequency of performing dominated actions (G1-dominated) have the same predictive power as task performance metrics. 


From the slopes ($m$) of the game theoretic metrics of G1-dominated (discarding a card that is known to be presently playable) and G3-dominant (playing a card that is known to be presently playable) actions, we see that even a slight change here correlates to large changes in human teamwork rating. When excluding RandomBot, a 1\% increase in the occurrence of G1-dominated actions causes a 9-point (26\%) drop in teamwork rating. Conversely, albeit less dramatically, we see a 1.21 point (3.5\%) increase in teamwork rating for every percent increase in G3-dominant frequency. Negative psychological bias~\cite{rozin2001negativity} may help explain the more pronounced effect of G1-dominated actions as these represent negative events in a team setting. 

The ``questionable moves'' data provides separate evidence for the negative effects of teammates' dominated actions. Both dominated moves are rare in the AI-only setting ($<1\%$ for G1-dominated other than RandomBot, $<0.5\%$ for G2-dominated other than OP+Aux). However, in human-AI play, 13\% of ``questionable move" button clicks that occurred after the AI discarded corresponded to G1-dominated actions, and 43\% of clicks after the AI played a card corresponded to G2-dominated actions. These data give strong evidence that humans react negatively when AI execute dominated actions. OP+Aux and AP+Aux's relatively high rate of playing G2-dominated actions may also help explain their low teamwork ratings (Figure~\ref{fig:hai_teamwork_ratings}) despite their best-in-pool inter-XP scores (Table~\ref{tab:ai_only_summary}).

The fact that task performance metrics are shown to have equal or lesser predictive power when compared to other objective measures of AI performance is a key result from our study which is explored more deeply in the following sections. %

\section{Discussion}
\label{sec:discussion}

Given the results presented throughout Sections~\ref{sec:ai_only_experiments} and \ref{sec:human_ai_experiments} let us revisit and assess the hypotheses we have presented throughout the paper. %
With the long-term goal of shaping RL reward functions for training human-collaborative agents, we first hypothesized that there exists one, or
several, objective metrics of AI behavior that are measurable in the absence of human interaction \emph{and} predictive of human preferences toward such AI once interaction commences in a human-AI team setting. %
Our results, particularly those summarized in Table~\ref{tab:regressions}, show that this hypothesis is supported, albeit with caveats. 
The objective metrics described in Table~\ref{tab:agent_metrics} are all measurable without human interaction (except human-AI game score, treated separately as a baseline), however they tend to have only weak-to-moderate predictive power of human preferences (\emph{r} = 0.2 to 0.5). 

The lack of objective metrics that are strongly predictive of subjective human preferences is of particular note given the narrow setting of Hanabi.
As was argued in Section~\ref{sec:introduction}, Hanabi acts as an ideal laboratory for investigating such correlations between AI behaviors and human reactions because of its highly-restrictive game mechanisms that eliminate many potentially confounding variables around communication. %
The fact that no particular metric of AI behavior was identified as being strongly predictive of human preferences in Hanabi seems to imply that it is unlikely that some form of ``general purpose'' reward function will be discovered to train human-collaborative RL agents for other human-AI teaming settings that are less constrained than Hanabi.

Such a result is important in the context of- and as a counterpoint to- other
reinforcement learning literature where it is often assumed---without any human testing to validate such assumption---AI that perform better in self-play or cross-play will be preferred by human teammates over AI that perform worse on those metrics~\cite{hu2020other,lucas2022any,strouse2021collaborating}; and thus you can simply design training reward functions that maximize these metrics. %
On the contrary, our results indicate that one should design a much more nuanced reward function that blends raw AI task performance (self-play, intra-XP, inter-XP) with information-theoretic (AD-entropy, ARD-entropy, CI, IC) and game-theoretic (G1-dominated, G2-dominated, G3-dominant) metrics; Section~\ref{subsec:sub_obj_metric_correlation}, and Table~\ref{tab:regressions} in particular, help quantify how such metrics might be weighted during training.

\subsection{Generalizations, Limitations, and Future Work}

How might these results extrapolate to other human-AI teaming tasks?
First we consider our information-theoretic metrics. %
While these metrics are some of the most strongly correlated to subjective teamwork ratings in Hanabi (see Table~\ref{tab:regressions}), it easy to imagine other teaming tasks that would not produce the same correlations. Take for example AD-entropy, a measure of the diversity of actions used by an AI (see Appendix~\ref{subapp:information_theoretic_metric_defs}). We postulate that AD-entropy correlates to a more favorable teamwork rating in Hanabi because humans may prefer AI partners that are more varied and explorative in their actions; they would not, for example, want a teammate who discards on every turn which would correlate to zero AD-entropy and would represent a very poor strategy in Hanabi. However, we could easily imagine other team tasks where humans would want highly repetitive and predictable teammates. In the extreme case, one could imagine a task where it is desirable for an AI teammate to always perform the exact same action or copy the action of a human; for example if the human-AI team were engaged in a manufacturing task. 

In contrast, we can make reasonable arguments as to why task performance and game-theoretic metrics may well correlate to human preferences toward AI teammates in many other settings beyond Hanabi. The concepts of \emph{payoffs} and \emph{dominated/dominant strategies} are broadly applicable to many decisions problems and games. While their numerical details may differ from one setting to another (e.g. payoff in Hanabi is 0 to 25 while in chess it may be a binary win-loss value), many of their characteristics and properties do not (e.g. higher payoffs are, by definition, more desirable than lower payoffs and payoffs can be increased by using non-dominated strategies)~\citep[Chp~4]{tadelis2013game}.

In order to avoid collecting personally-identifiable information in this study, we collected only demographic data that was relevant to subjects' Hanabi-playing experience, and do not have data such as age, AI background, and cultural background, so we can only infer this based on the recruitment process. It is possible that there is an age bias towards people in the 18-25 year old range given the university setting, but the MIT BRL subject pool, from which we obtained the bulk of our participants, has a much broader demographic, more reflective of the Cambridge, MA population. This is a population for which we expect some preexisting exposure to AI systems through common consumer products, but not one in which we would expect a preponderance of specific AI expertise. It is possible that participants from particular demographics may have different reactions to AI agents, and this remains a question our work is unable to answer precisely.


We make several hypotheses for future experiments. First: humans with sufficient experience in a team task are able to intuit dominated and dominant strategies within that task; at least within short time horizons due to humans' limited capacity for \emph{bounded rationality} \cite{simon1990bounded}. 
Second: humans will prefer teammates that both maximize the team payoff (i.e. task performance) while avoiding use of any strategy that may be \emph{perceived} to be dominated, and thus human-collaborative AI should be trained with rewards shaped to account for both of these objectives where data from Table~\ref{tab:regressions} can be used to inform the relative weighting of each. %
Maximizing payoff and avoiding dominated strategies are two sides of the same coin. However, given a human's limited ability to predict over long sequences of decisions \cite{simon1990bounded}, an AI teammate could utilize non-dominated strategies that have high potential payoff, but that appear to be dominated in the short term. Examples of this can already be found in human-AI \emph{competitive} games such as the famed ``Move 37'' in the Go match between Lee Sedol and AlphaGo, which was initially perceived to be a mistake (e.g. dominated action) made by AlphaGo only later to be understood as the pivotal move that allowed AlphaGo to win the game \cite{holcomb2018overview}. %
Within collaborative games, this \emph{appearance} of dominated actions may lead to trust-breaking with the human counterpart, thus, should be trained out of the AI in order to maximize subjective preferences toward the AI, or the human should be informed of them beforehand along with their reasoning \cite{leahy2024tell}. 

\section{Conclusion}
\label{sec:conclusion}

Despite the historical focus on task scores in AI development and reinforcement learning in particular, we show that high scores do not tell the whole story about human-AI teams. Although Hanabi is an extreme example of teaming, we believe that these results extend to the broader discussion of human acceptance of AI systems. We confirm the results of previous human experiments in Hanabi, and expand analysis to more agent types, humans, and metrics, finding several objective metrics of AI behavior that are comparable-or-better predictors of teaming preference than the final score achieved by the team. Among these metrics, game theoretic strategic dominance is a promising avenue of exploration for preemptively avoiding human rejection when designing learning-based AI teammates for humans.

\section*{Impact Statement}


We believe that our results---particularly, the quantification of how AI behavioral metrics correlate with human preferences toward teammates---can have broad impact on the development of human-collaborative AI agents. We expect that the behavioral metrics we've identified can and will be used to better shape reward functions for training agents using reinforcement learning that are more human-acceptable than those trained with a myopic focus on raw task performance.
Negative societal impacts could occur, however, if better collaborative agents were developed for malicious purposes, or if these metrics were directly attacked by an adversary to produce high-performing agents that are nonetheless rejected by its users.

\nocite{langley00}

\bibliography{bibliography}

\begin{thebibliography}{51}
\providecommand{\natexlab}[1]{#1}
\providecommand{\url}[1]{\texttt{#1}}
\expandafter\ifx\csname urlstyle\endcsname\relax
  \providecommand{\doi}[1]{doi: #1}\else
  \providecommand{\doi}{doi: \begingroup \urlstyle{rm}\Url}\fi

\bibitem[Bansal et~al.(2017)Bansal, Pachocki, Sidor, Sutskever, and Mordatch]{bansal2017emergent}
Bansal, T., Pachocki, J., Sidor, S., Sutskever, I., and Mordatch, I.
\newblock Emergent complexity via multi-agent competition.
\newblock \emph{arXiv preprint arXiv:1710.03748}, 2017.

\bibitem[Bard et~al.(2020)Bard, Foerster, Chandar, Burch, Lanctot, Song, Parisotto, Dumoulin, Moitra, Hughes, et~al.]{bard2020hanabi}
Bard, N., Foerster, J.~N., Chandar, S., Burch, N., Lanctot, M., Song, H.~F., Parisotto, E., Dumoulin, V., Moitra, S., Hughes, E., et~al.
\newblock The hanabi challenge: A new frontier for ai research.
\newblock \emph{Artificial Intelligence}, 280:\penalty0 103216, 2020.

\bibitem[Berner et~al.(2019)Berner, Brockman, Chan, Cheung, Dkebiak, Dennison, Farhi, Fischer, Hashme, Hesse, et~al.]{berner2019dota}
Berner, C., Brockman, G., Chan, B., Cheung, V., Dkebiak, P., Dennison, C., Farhi, D., Fischer, Q., Hashme, S., Hesse, C., et~al.
\newblock Dota 2 with large scale deep reinforcement learning.
\newblock \emph{arXiv preprint arXiv:1912.06680}, 2019.

\bibitem[Bogin et~al.(2018)Bogin, Geva, and Berant]{bogin2018emergence}
Bogin, B., Geva, M., and Berant, J.
\newblock Emergence of communication in an interactive world with consistent speakers.
\newblock \emph{arXiv preprint arXiv:1809.00549}, 2018.

\bibitem[Canaan et~al.(2020)Canaan, Gao, Togelius, Nealen, and Menzel]{canaan2020generating}
Canaan, R., Gao, X., Togelius, J., Nealen, A., and Menzel, S.
\newblock Generating and adapting to diverse ad-hoc cooperation agents in hanabi.
\newblock \emph{arXiv preprint arXiv:2004.13710}, 2020.

\bibitem[Carroll et~al.(2019)Carroll, Shah, Ho, Griffiths, Seshia, Abbeel, and Dragan]{carroll2019utility}
Carroll, M., Shah, R., Ho, M.~K., Griffiths, T., Seshia, S., Abbeel, P., and Dragan, A.
\newblock On the utility of learning about humans for human-ai coordination.
\newblock \emph{Advances in Neural Information Processing Systems}, 32, 2019.

\bibitem[Christiano et~al.(2017)Christiano, Leike, Brown, Martic, Legg, and Amodei]{christiano2017deep}
Christiano, P.~F., Leike, J., Brown, T., Martic, M., Legg, S., and Amodei, D.
\newblock Deep reinforcement learning from human preferences.
\newblock \emph{Advances in neural information processing systems}, 30, 2017.

\bibitem[Cooke \& Shope(2004)Cooke and Shope]{cooke2004synthetic}
Cooke, N.~J. and Shope, S.~M.
\newblock Synthetic task environments for teams: Certt’s uav-ste.
\newblock In \emph{Handbook of human factors and ergonomics methods}, pp.\  476--483. CRC Press, Boca Raton, FL, USA, 2004.

\bibitem[Cover \& Thomas(2006)Cover and Thomas]{cover2006elements}
Cover, T.~M. and Thomas, J.~A.
\newblock \emph{Elements of Information Theory Second Edition}.
\newblock Wiley-Interscience, USA, 2006.
\newblock ISBN 0471241954.

\bibitem[(FAIR)† et~al.(2022)(FAIR)†, Bakhtin, Brown, Dinan, Farina, Flaherty, Fried, Goff, Gray, Hu, et~al.]{meta2022human}
(FAIR)†, M. F. A. R. D.~T., Bakhtin, A., Brown, N., Dinan, E., Farina, G., Flaherty, C., Fried, D., Goff, A., Gray, J., Hu, H., et~al.
\newblock Human-level play in the game of diplomacy by combining language models with strategic reasoning.
\newblock \emph{Science}, 378\penalty0 (6624):\penalty0 1067--1074, 2022.

\bibitem[Gray et~al.(2020)Gray, Lerer, Bakhtin, and Brown]{gray2020human}
Gray, J., Lerer, A., Bakhtin, A., and Brown, N.
\newblock Human-level performance in no-press diplomacy via equilibrium search.
\newblock \emph{arXiv preprint arXiv:2010.02923}, 2020.

\bibitem[Guznov et~al.(2020)Guznov, Lyons, Pfahler, Heironimus, Woolley, Friedman, and Neimeier]{guznov2020robot}
Guznov, S., Lyons, J., Pfahler, M., Heironimus, A., Woolley, M., Friedman, J., and Neimeier, A.
\newblock Robot transparency and team orientation effects on human--robot teaming.
\newblock \emph{International Journal of Human--Computer Interaction}, 36\penalty0 (7):\penalty0 650--660, 2020.

\bibitem[Hoff \& Bashir(2015)Hoff and Bashir]{hoff2015trust}
Hoff, K.~A. and Bashir, M.
\newblock Trust in automation: Integrating empirical evidence on factors that influence trust.
\newblock \emph{Human factors}, 57\penalty0 (3):\penalty0 407--434, 2015.

\bibitem[Hoffman(2019)]{hoffman2019evaluating}
Hoffman, G.
\newblock Evaluating fluency in human--robot collaboration.
\newblock \emph{IEEE Transactions on Human-Machine Systems}, 49\penalty0 (3):\penalty0 209--218, 2019.

\bibitem[Holcomb et~al.(2018)Holcomb, Porter, Ault, Mao, and Wang]{holcomb2018overview}
Holcomb, S.~D., Porter, W.~K., Ault, S.~V., Mao, G., and Wang, J.
\newblock Overview on deepmind and its alphago zero ai.
\newblock In \emph{Proceedings of the 2018 international conference on big data and education}, pp.\  67--71, 2018.

\bibitem[Hu \& Foerster(2019)Hu and Foerster]{hu2019simplified}
Hu, H. and Foerster, J.~N.
\newblock Simplified action decoder for deep multi-agent reinforcement learning.
\newblock \emph{arXiv preprint arXiv:1912.02288}, 2019.

\bibitem[Hu et~al.(2020)Hu, Lerer, Peysakhovich, and Foerster]{hu2020other}
Hu, H., Lerer, A., Peysakhovich, A., and Foerster, J.
\newblock “other-play” for zero-shot coordination.
\newblock In \emph{International Conference on Machine Learning}, pp.\  4399--4410. PMLR, 2020.

\bibitem[Hu et~al.(2021)Hu, Lerer, Cui, Pineda, Brown, and Foerster]{hu2021off}
Hu, H., Lerer, A., Cui, B., Pineda, L., Brown, N., and Foerster, J.
\newblock Off-belief learning.
\newblock In \emph{International Conference on Machine Learning}, pp.\  4369--4379. PMLR, 2021.

\bibitem[Hu et~al.(2022)Hu, Wu, Lerer, Foerster, and Brown]{hu2022human}
Hu, H., Wu, D.~J., Lerer, A., Foerster, J., and Brown, N.
\newblock Human-ai coordination via human-regularized search and learning.
\newblock \emph{arXiv preprint arXiv:2210.05125}, 2022.

\bibitem[Jaques et~al.(2019)Jaques, Lazaridou, Hughes, Gulcehre, Ortega, Strouse, Leibo, and De~Freitas]{jaques2019social}
Jaques, N., Lazaridou, A., Hughes, E., Gulcehre, C., Ortega, P., Strouse, D., Leibo, J.~Z., and De~Freitas, N.
\newblock Social influence as intrinsic motivation for multi-agent deep reinforcement learning.
\newblock In \emph{International Conference on Machine Learning}, pp.\  3040--3049. PMLR, 2019.

\bibitem[Kaufmann et~al.(2023)Kaufmann, Bauersfeld, Loquercio, M{\"u}ller, Koltun, and Scaramuzza]{kaufmann2023champion}
Kaufmann, E., Bauersfeld, L., Loquercio, A., M{\"u}ller, M., Koltun, V., and Scaramuzza, D.
\newblock Champion-level drone racing using deep reinforcement learning.
\newblock \emph{Nature}, 620\penalty0 (7976):\penalty0 982--987, 2023.

\bibitem[Kiran et~al.(2021)Kiran, Sobh, Talpaert, Mannion, Al~Sallab, Yogamani, and P{\'e}rez]{kiran2021deep}
Kiran, B.~R., Sobh, I., Talpaert, V., Mannion, P., Al~Sallab, A.~A., Yogamani, S., and P{\'e}rez, P.
\newblock Deep reinforcement learning for autonomous driving: A survey.
\newblock \emph{IEEE Transactions on Intelligent Transportation Systems}, 23\penalty0 (6):\penalty0 4909--4926, 2021.

\bibitem[Kurach et~al.(2020)Kurach, Raichuk, Stanczyk, Zajkac, Bachem, Espeholt, Riquelme, Vincent, Michalski, Bousquet, et~al.]{kurach2020google}
Kurach, K., Raichuk, A., Stanczyk, P., Zajkac, M., Bachem, O., Espeholt, L., Riquelme, C., Vincent, D., Michalski, M., Bousquet, O., et~al.
\newblock Google research football: A novel reinforcement learning environment.
\newblock In \emph{Proceedings of the AAAI conference on artificial intelligence}, volume 34:04, pp.\  4501--4510, 2020.

\bibitem[Leahy \& Siu(2024)Leahy and Siu]{leahy2024tell}
Leahy, K. and Siu, H.~C.
\newblock Tell me what you want (what you really, really want): Addressing the expectation gap for goal conveyance from humans to robots.
\newblock \emph{arXiv preprint arXiv:2403.14344}, 2024.

\bibitem[Lerer et~al.(2020)Lerer, Hu, Foerster, and Brown]{lerer2020improving}
Lerer, A., Hu, H., Foerster, J., and Brown, N.
\newblock Improving policies via search in cooperative partially observable games.
\newblock In \emph{Proceedings of the AAAI Conference on Artificial Intelligence}, volume 34:05, pp.\  7187--7194, 2020.

\bibitem[Lin et~al.(2011)Lin, Foster, and Ungar]{lin2011vif}
Lin, D., Foster, D.~P., and Ungar, L.~H.
\newblock Vif regression: a fast regression algorithm for large data.
\newblock \emph{Journal of the American Statistical Association}, 106\penalty0 (493):\penalty0 232--247, 2011.

\bibitem[Lowe et~al.(2019)Lowe, Foerster, Boureau, Pineau, and Dauphin]{lowe2019pitfalls}
Lowe, R., Foerster, J., Boureau, Y.-L., Pineau, J., and Dauphin, Y.
\newblock On the pitfalls of measuring emergent communication.
\newblock \emph{arXiv preprint arXiv:1903.05168}, 2019.

\bibitem[Lucas \& Allen(2022)Lucas and Allen]{lucas2022any}
Lucas, K. and Allen, R.~E.
\newblock Any-play: An intrinsic augmentation for zero-shot coordination.
\newblock In \emph{Proceedings of the 21st International Conference on Autonomous Agents and Multiagent Systems}, pp.\  853--861, 2022.

\bibitem[Lupu et~al.(2021)Lupu, Cui, Hu, and Foerster]{lupu2021trajectory}
Lupu, A., Cui, B., Hu, H., and Foerster, J.
\newblock Trajectory diversity for zero-shot coordination.
\newblock In \emph{International conference on machine learning}, pp.\  7204--7213. PMLR, 2021.

\bibitem[Mnih et~al.(2013)Mnih, Kavukcuoglu, Silver, Graves, Antonoglou, Wierstra, and Riedmiller]{mnih2013playing}
Mnih, V., Kavukcuoglu, K., Silver, D., Graves, A., Antonoglou, I., Wierstra, D., and Riedmiller, M.
\newblock Playing atari with deep reinforcement learning.
\newblock \emph{arXiv preprint arXiv:1312.5602}, 2013.

\bibitem[Nekoei et~al.(2021)Nekoei, Badrinaaraayanan, Courville, and Chandar]{nekoei2021continuous}
Nekoei, H., Badrinaaraayanan, A., Courville, A., and Chandar, S.
\newblock Continuous coordination as a realistic scenario for lifelong learning.
\newblock In \emph{International Conference on Machine Learning}, pp.\  8016--8024. PMLR, 2021.

\bibitem[O'Dwyer(2019)]{odwyer2019hanabi}
O'Dwyer, A.
\newblock quuxplusone/hanabi: framework for writing bots that play hanabi.
\newblock https://github.com/Quuxplusone/Hanabi/, 2019.

\bibitem[O’Neill et~al.(2022)O’Neill, McNeese, Barron, and Schelble]{oneill2022human}
O’Neill, T., McNeese, N., Barron, A., and Schelble, B.
\newblock Human--autonomy teaming: A review and analysis of the empirical literature.
\newblock \emph{Human factors}, 64\penalty0 (5):\penalty0 904--938, 2022.

\bibitem[Rafailov et~al.(2024)Rafailov, Sharma, Mitchell, Manning, Ermon, and Finn]{rafailov2024direct}
Rafailov, R., Sharma, A., Mitchell, E., Manning, C.~D., Ermon, S., and Finn, C.
\newblock Direct preference optimization: Your language model is secretly a reward model.
\newblock \emph{Advances in Neural Information Processing Systems}, 36, 2024.

\bibitem[Rheu et~al.(2021)Rheu, Shin, Peng, and Huh-Yoo]{rheu2021systematic}
Rheu, M., Shin, J.~Y., Peng, W., and Huh-Yoo, J.
\newblock Systematic review: Trust-building factors and implications for conversational agent design.
\newblock \emph{International Journal of Human--Computer Interaction}, 37\penalty0 (1):\penalty0 81--96, 2021.

\bibitem[Richens et~al.(2020)Richens, Lee, and Johri]{richens2020improving}
Richens, J.~G., Lee, C.~M., and Johri, S.
\newblock Improving the accuracy of medical diagnosis with causal machine learning.
\newblock \emph{Nature communications}, 11\penalty0 (1):\penalty0 3923, 2020.

\bibitem[Rozin \& Royzman(2001)Rozin and Royzman]{rozin2001negativity}
Rozin, P. and Royzman, E.~B.
\newblock Negativity bias, negativity dominance, and contagion.
\newblock \emph{Personality and social psychology review}, 5\penalty0 (4):\penalty0 296--320, 2001.

\bibitem[Shah et~al.(2018)Shah, Kamath, Shah, and Li]{shah2018bayesian}
Shah, A., Kamath, P., Shah, J.~A., and Li, S.
\newblock Bayesian inference of temporal task specifications from demonstrations.
\newblock \emph{Advances in Neural Information Processing Systems}, 31, 2018.

\bibitem[Silver et~al.(2018)Silver, Hubert, Schrittwieser, Antonoglou, Lai, Guez, Lanctot, Sifre, Kumaran, Graepel, et~al.]{silver2018general}
Silver, D., Hubert, T., Schrittwieser, J., Antonoglou, I., Lai, M., Guez, A., Lanctot, M., Sifre, L., Kumaran, D., Graepel, T., et~al.
\newblock A general reinforcement learning algorithm that masters chess, shogi, and go through self-play.
\newblock \emph{Science}, 362\penalty0 (6419):\penalty0 1140--1144, 2018.

\bibitem[Simon(1990)]{simon1990bounded}
Simon, H.~A.
\newblock Bounded rationality.
\newblock \emph{Utility and probability}, pp.\  15--18, 1990.

\bibitem[Siu et~al.(2021)Siu, Pe{\~n}a, Chen, Zhou, Lopez, Palko, Chang, and Allen]{siu2021evaluation}
Siu, H.~C., Pe{\~n}a, J., Chen, E., Zhou, Y., Lopez, V., Palko, K., Chang, K., and Allen, R.~E.
\newblock Evaluation of human-ai teams for learned and rule-based agents in hanabi.
\newblock \emph{Advances in Neural Information Processing Systems}, 34, 2021.

\bibitem[Strouse et~al.(2021)Strouse, McKee, Botvinick, Hughes, and Everett]{strouse2021collaborating}
Strouse, D., McKee, K., Botvinick, M., Hughes, E., and Everett, R.
\newblock Collaborating with humans without human data.
\newblock \emph{Advances in Neural Information Processing Systems}, 34:\penalty0 14502--14515, 2021.

\bibitem[Sutton \& Barto(2018)Sutton and Barto]{sutton2018reinforcement}
Sutton, R.~S. and Barto, A.~G.
\newblock \emph{Reinforcement learning: An introduction}.
\newblock MIT Press, Cambridge, MA, USA, 2018.

\bibitem[Suzuki(2017)]{suzuki2017overview}
Suzuki, K.
\newblock Overview of deep learning in medical imaging.
\newblock \emph{Radiological physics and technology}, 10\penalty0 (3):\penalty0 257--273, 2017.

\bibitem[Tadelis(2013)]{tadelis2013game}
Tadelis, S.
\newblock \emph{Game theory: an introduction}.
\newblock Princeton university press, USA, 2013.

\bibitem[Tauer \& Harackiewicz(1999)Tauer and Harackiewicz]{tauer1999winning}
Tauer, J.~M. and Harackiewicz, J.~M.
\newblock Winning isn't everything: Competition, achievement orientation, and intrinsic motivation.
\newblock \emph{Journal of Experimental Social Psychology}, 35\penalty0 (3):\penalty0 209--238, 1999.

\bibitem[Thaler \& Sunstein(2009)Thaler and Sunstein]{thaler2009nudge}
Thaler, R.~H. and Sunstein, C.~R.
\newblock \emph{Nudge: Improving Decisions About Health, Wealth, and Happiness}.
\newblock Penguin, 2009.

\bibitem[Vinyals et~al.(2019)Vinyals, Babuschkin, Czarnecki, Mathieu, Dudzik, Chung, Choi, Powell, Ewalds, Georgiev, et~al.]{vinyals2019grandmaster}
Vinyals, O., Babuschkin, I., Czarnecki, W.~M., Mathieu, M., Dudzik, A., Chung, J., Choi, D.~H., Powell, R., Ewalds, T., Georgiev, P., et~al.
\newblock Grandmaster level in starcraft ii using multi-agent reinforcement learning.
\newblock \emph{Nature}, 575\penalty0 (7782):\penalty0 350--354, 2019.

\bibitem[Wurman et~al.(2022)Wurman, Barrett, Kawamoto, MacGlashan, Subramanian, Walsh, Capobianco, Devlic, Eckert, Fuchs, et~al.]{wurman2022outracing}
Wurman, P.~R., Barrett, S., Kawamoto, K., MacGlashan, J., Subramanian, K., Walsh, T.~J., Capobianco, R., Devlic, A., Eckert, F., Fuchs, F., et~al.
\newblock Outracing champion gran turismo drivers with deep reinforcement learning.
\newblock \emph{Nature}, 602\penalty0 (7896):\penalty0 223--228, 2022.

\bibitem[Zhao et~al.(2020)Zhao, Borovikov, de~Mesentier~Silva, Beirami, Rupert, Somers, Harder, Kolen, Pinto, Pourabolghasem, et~al.]{zhao2020winning}
Zhao, Y., Borovikov, I., de~Mesentier~Silva, F., Beirami, A., Rupert, J., Somers, C., Harder, J., Kolen, J., Pinto, J., Pourabolghasem, R., et~al.
\newblock Winning is not everything: Enhancing game development with intelligent agents.
\newblock \emph{IEEE Transactions on Games}, 12\penalty0 (2):\penalty0 199--212, 2020.

\bibitem[Ziegler et~al.(2019)Ziegler, Stiennon, Wu, Brown, Radford, Amodei, Christiano, and Irving]{ziegler2019fine}
Ziegler, D.~M., Stiennon, N., Wu, J., Brown, T.~B., Radford, A., Amodei, D., Christiano, P., and Irving, G.
\newblock Fine-tuning language models from human preferences.
\newblock \emph{arXiv preprint arXiv:1909.08593}, 2019.

\end{thebibliography}
\bibliographystyle{icml2025}

\newpage
\appendix
\onecolumn

\section{Hanabi Game Rules}
\label{app:hanabi_rules}

A team of players works together to stack cards in five different piles (referred to as ``fireworks''), one for each suit (color) and by ascending rank (number). 
While Hanabi can be played with 2-5 players, this work exclusively considers two-player games (either parings of AI-AI or human-AI) where the deck is composed of 50 cards, five suits, each suit having three 1s, two 2s, two 3s, two 4s, and one 5, and each player is dealt a hand of five cards.

Hanabi games start with eight \emph{hint tokens} and three \emph{bomb tokens} (strikes are referred to as ``bombs''). Each turn, a player may \emph{discard} a card from their hand, \emph{play} a card from their hand, or give a \emph{hint} to a teammate. Whenever a card is played or discarded, it is revealed and a new card is drawn from the deck. A correctly-played card is placed in the appropriate firework; an incorrectly-played card is discarded and the team loses one bomb. 
The game ends when all fireworks have been completed, the deck is empty (and each player has one additional turn), or the team loses 3 bombs.
The final score is the sum of the top card in each firework, for a maximum of 25 points.

Hanabi’s difficulty lies in the fact that players can only see their teammate’s hand---never their own---and communication about cards is strictly limited. %
Communication comes in the form of \emph{hinting} whereby one player can user their turn to reveal the suit or rank of cards in another player's hand. The number of hints is limited by ``hint tokens'' that are expended upon hinting, but can be earned back when piles are completed or cards are discarded. 
Implied communication may arise from a player's choice to play (i.e. stack) or discard a card, but even correctly inferring such actions as subtle communication---rather than simply a means to immediately score points, or even a mistake of the player---is a challenge for the rest of the team. 
There is no overt communication allowed---whether it be verbal, text, or body language---that would help players coordinate on strategies or intents.
While communication is undoubtedly an essential component of coordination in most real-world teaming tasks, the restriction of communication in Hanabi allows us to eliminate many variables that may confound the testing of our hypotheses. %
    
Hanabi is a \textit{purely cooperative} game with imperfect information, and limited, codified communication. These properties make it an interesting challenge for teaming since players must consider the reasons for their teammates' actions and any implied information, while avoiding misinterpretations. Bard et al. present a more complete treatment of Hanabi and its properties as an AI problem \cite{bard2020hanabi}.

\section{Agents Informed by Human Data}
\label{app:human_data_agents}

As noted in Section \ref{subsec:ai-only-evaluations}, agents in the literature that included the use of human data could not be evaluated, as they were not released, despite the fact that the ones that used human data as a regularizer for reinforcement learning (piKL3) reported some of the best performance when with paired with humans \cite{hu2022human}. The authors cited difficulties regarding purchased human data. It is unclear where they would be in the space of our AI-only metrics, though it is likely that they would be closer to where an experienced human player would be.

We also note, however, that a weakness in agents trained on human data is that they likely do not capture certain elements of playing with a human well. They likely do not, for example, capture the impact of loss of trust during gameplay, and how it may cause a shift in human effort or strategy. Additionally, training on human-only games likely does not explore some significant portions of the state space, as all players are starting with some base level of reasoning. 

\section{Objective Metrics Selection}
\label{app:objective_metrics_selection}

Here we present rigorous definitions for the objective metrics presented in Section~\ref{subsec:results_objective_metrics_selection}

\subsection{Task Performance}
\label{subapp:task_performance_objective_metrics_selection}



Here we present mathematical definitions for the task performance metrics of self-play, intra-algorithm cross-play (intra-XP), and inter-algorithm cross-play (inter-XP) scores. These definitions are taken, almost verbatim, from ~\cite{lucas2022any} but are reproduced here for completeness.

Let us consider two agents (whether they be AI or humans) playing as teammates in a two-player game of Hanabi; let us represent them as $i$ and $j$. The \emph{policies} of the respective agents---i.e. the mechanism by which a player maps its current observation of the game state to its next action---are represented as $\pi_i$ and $\pi_j$. The final game score---referred to more generally as the \emph{return}---for a particular instance of this two-player game is represented as $G$. The expected score (expected return) of such games is represented as $V\left(\pi_1, \pi_j \right) = \mathbb{E}\left[ G \right]$.
Let $\alpha$ represent a stochastic \emph{policy-generator algorithm}---such as a reinforcement learning algorithm---that generates a policy $\pi$ from a distribution of possible policies $\Pi_\alpha$. %
Let $\mathcal{C}$ represent the set of all policy generator algorithms such that $\alpha \in \mathcal{C}$. %

Now we can define a set of evaluation metrics based on expected game scores.

\textbf{Self-Play} (SP) scores %
represent the expected game score of any agent $i$ generated (i.e. trained) by algorithm $\alpha$ when playing cooperatively with a copy of itself.
\begin{equation}\label{eq:selfplay}
    J_{\text{SP}}(\alpha) = \mathbb{E}_{\pi_{i} \sim \Pi_\alpha}\left[V(\pi_i,\pi_i)\right]
\end{equation} 

\textbf{Intra-Algorithm Cross-Play} (intra-XP) score represents the expected game score of any agent $i$ generated by algorithm $\alpha$ when partnered with any agent $j$ that has been generated \emph{independently} by the same algorithm $\alpha$.
\begin{equation}\label{eq:intraalg}
    J_{\text{intra-XP}}(\alpha) = \mathbb{E}_{\pi_{i} \sim \Pi_\alpha, \pi_{j} \sim \Pi_\alpha}\left[V(\pi_i,\pi_j)\right]
\end{equation} %

\textbf{Inter-Algorithm Cross-Play} (inter-XP) %
score is the expected return of any agent $i$ generated from algorithm $\alpha \in \mathcal{C}$ when partnered with any other agent $j$ generated independently from a separate algorithm $\beta \in \mathcal{C}\setminus \alpha$
\begin{equation}\label{eq:interalg}
    J_{\text{inter-XP}}(\alpha) = \mathbb{E}_{\pi_{i} \sim \Pi_{\alpha \in \mathcal{C}}, \pi_{j} \sim \Pi_{\beta \in \mathcal{C}\setminus\alpha}}\left[V(\pi_i,\pi_j)\right]
\end{equation} %

\subsection{Information-Theoretic}
\label{subapp:information_theoretic_metric_defs}

These metrics center around the distributions of agents' actions over the course of many games.
Action distribution is determined by counts of agent actions from game traces, and action-response is a matrix version of the same that considers pairs of actions---what one's teammate did and what one did in response. Despite being computationally similar, the entropies of these distributions represent different metrics and the latter contains more granular information that is also more dependent on partner behavior. For these entropies, we consider the Shannon entropy~\cite[Chp~2]{cover2006elements} of each distribution,

\begin{equation}
    H(X) = -\sum_{x\in X} p(x) log(p(x)),
\end{equation}

\noindent where $x$ is a distinct action in the set of $X$ possible actions ($|X| = 20$) for the case of action entropy (AD-entropy), and $x$ is a distinct combination of action and response in the set of all $X$ possible action-response pairs ($|X| = 20\times 20 = 400$) for the case of action-response entropy (ARD-entropy).

Instantaneous coordination is a measure of the mutual information between pairs of subsequent actions \cite{jaques2019social} in a game trajectory, defined as

\begin{equation}\label{eqn:instant_coord_def}
    IC = I(a_t^A; a_{t+1}^B)
\end{equation}

\noindent where $a_t^A$ is the action taken by agent $A$ at time $t$ and $a_{t+1}^B$ is the action taken by agent $B$ at time $t+1$. Jacques et al. define IC separately for agent $A$'s \emph{actions} and \emph{messages} at time $t$, influencing agent $B$ \cite{jaques2019social}. However, given the potential overlap between acting and communicating in Hanabi, we consider all Hanabi moves (hint, discard, play) as actions, collapsing the two IC definitions given by Jaques et al. \cite{jaques2019social} into the single definition given in Equation~\ref{eqn:instant_coord_def}.

Context independence (CI) comes from the emergent communications literature in reinforcement learning \cite{bogin2018emergence,lowe2019pitfalls}, and is defined as 

\begin{equation}
    CI(p_{cm}, p_{mc}) = \frac{1}{|C|}\sum_c p_{mc}(m^c|c) p_{cm}(c|m^c),
\end{equation}

\noindent where $m$ is a message, $c$ is a concept, $m^c$ is the most likely message given a concept $c$, $|c|$ is the space of possible concepts, $p_{cm} (c|m)$ is the conditional
probability of a concept given a message, and similarly for $p_{mc}(m|c)$.

A difference between Hanabi and settings common to emergent communications literature is that the latter typically assumes a clear distinction between \emph{actions} and \emph{messages}, and where messages occur over a no-cost ``cheap talk'' channel. Hanabi has clearly-defined non-zero costs for communication, and less of a distinction between actions and messages, since even non-hinting actions usually reveal information, either explicitly about the state of the game, or implicitly about the behavior of the player.

The use of the context independence metric \cite{bogin2018emergence} requires some special considerations for Hanabi, especially for two issues:

For context semantics, CI was originally developed with the framework of a simple referential game, with specific, well-defined \emph{concepts}, and asked the question of whether a \emph{message} was sent due to the presence of a concept in the world. Extracting semantics from the observation space of Hanabi requires some parsing to understand, for example whether there were a particular number of hint tokens left, or if the card that was just played had a particular attribute. Furthermore, concepts used in Hanabi conventions (among humans) typically involve several attributes, such as hinting a newly-drawn, playable card as an indicator that it should be played immediately, even without full information (player action, card age, card playability, hint age), and discarding the oldest card without any direct hints on it (player action, card age, hint status)

To deal with this issue, we took inspiration from Shah et al. \cite{shah2018bayesian} and their method of representing task states. We constructed a set of primitive concepts including both instantaneous (e.g. number of hint tokens, color of card) and contextual  (e.g. card was just hinted) information, and sampled atomic and multi-part logical predicates. Atomic predicates were expressed as first-order logical formulae, and multi-part predicates allowed for the use of \emph{AND} and \emph{OR} operators between predicates, and potentially \emph{NOT} operators before atomic propositions, thus forming second-order logical formulae. \emph{Concepts} in the CI usage were then defined to be first- or second-order formulae describing the state of the game, while \emph{messages} in the CI usage were the associated player actions taken during the presence of a concept. In practice, the potential formula space is arbitrarily large, so for our calculations, we sampled 500 formulas up to a tree depth of 3. 

A further issue comes in the interpretation of the CI score. The original usage was such that higher context-independence was an indication of emergent communication, with the intuition being that a concept was consistently referred to by a message, regardless of other distractor concepts. A high CI indicates that vocabulary usage to describe concepts was closer to one-to-one. This intuition holds in an environment where the number of potential messages is equal to or greater than the number of concepts.

In Hanabi, we quickly see that the reverse is true: the number of concepts far exceeds the number of potential actions (play, discard, hint color, hint rank). Hanabi is structured as a game of \emph{implication}, with the small number of player actions communicating more information than is possible explicitly, with the precise meaning of the action dependent on the state of the game. Achieving high levels of communication in this situation thus requires context \emph{dependence}, so the usual interpretation of CI is reversed: a \emph{lower} value is a better indicator of emergent communication.

\subsection{Game-Theoretic}
\label{subapp:game_theoretic_metric_defs}

Dominated/dominant move frequency are a pair of game theoretic metrics that attempt to formalize the identification of the kind of ``irrational behaviors'' that caused particularly strong human reactions in previous work \cite{siu2021evaluation}. All move frequency metrics are normalized on a per-game basis, according to:

\begin{equation}
    F_{move} = \sum_{i=1}^{n} \frac{t_{move,i}}{t_{game,i}},
\end{equation}

\noindent where $t_{move,i}$ is the number of turns in game $i$ in which an agent in question makes a particular type of move (outlined below) $t_{game,i}$ is the number of turns taken by the same agent in game $i$, and $n$ is the number of games under consideration.

We first determine a set of single-step dominated and dominant strategies for the Hanabi domain. A dominated strategy in this cooperative context is one where a superior alternative exists, and a dominant strategy is one where no superior alternative exists. An advantage in our human-AI context is that since we are concerned with \textit{human} reactions, we do not need to consider the entire game tree, but rather only a few moves ahead, limited by humans' \textit{bounded rationality} \cite{simon1990bounded}. We consider the following set of moves:

\begin{itemize}
    \item \textbf{G1-dominated: Discard playable.} Discarding a card that is known to be \textit{presently playable}. This is a \emph{single-step strictly dominated} strategy because there is always an option to play any card in one's hand, which in the case of full knowledge of playability, is strictly superior to discarding.
    
    \item \textbf{G2-dominated: Play unplayable.} Playing a card that is known to be \textit{presently unplayable}. This is also a \emph{single-step strictly dominated} strategy because it leads to a negative outcome (causing a strike, potentially leading to a loss), and there always exist other move options (i.e. one is never forced into such a move).
    
    \item \textbf{G3-dominant: Play playable.} Playing a card that is known to be \textit{presently playable}. This is a \emph{single-step weakly dominant} strategy because it provides an immediate +1 reward, but there may be other moves that are equivalent or superior for longer-term reward.

\end{itemize}

These strategies can be identified in a single turn, so they are well within a human's bounded rationality horizon, though such a restriction is conservative and not strictly necessary. We hypothesize that the use of dominated strategies is negatively correlated with human perceptions of teamwork and vice versa for dominant strategies.

\subsection{Excluded Metrics} %
\label{subapp:excluded_metrics}
We considered, but ultimately, omitted, other emergent communication metrics, such as causal influence of communication~\cite{lowe2019pitfalls}, social influence~\cite{jaques2019social}, and speaker consistency~\cite{jaques2019social}. These metrics use counterfactual play (i.e. ``imagining'' other ways the game could have played out) and cheap talk (i.e. direct communication of game information); and thus would require modification to the Hanabi environment. 

We also considered the use of information theoretic divergence metrics---comparisons of action and action-response distributions of one agent against another. However, this choice would have necessitated all pairwise comparisons of agents, or comparison against a ``reference'' agent, which would require further selection and justification.

\section{Data Analysis}
\label{subapp:data_analysis}

We evaluated the collection of hypotheses that there is a correlation between each of the objective AI-only metrics (our independent variables) against subjective human preferences regarding teaming with the associated agents (our dependent variables). Evaluations of post-block sentiment used a sum of the responses B3 to B8 in Table \ref{tab:post-bot-left-experiment-right}-left as a single Likert scale (the \emph{teamwork rating}). Post-experiment sentiment analysis (Supplemental Material \ref{appendix:post_game_comparisons}) took the sum of P1 to P7 in Table \ref{tab:post-bot-left-experiment-right}-right. 

There is significant correlation between many of the independent variables---an unavoidable feature of the experiment since the agents were chosen from the literature rather than designed specifically for this experiment. As this multicollinearity would cause problems for a multiple regression analysis, we opt to simply perform single-variable regressions against each metric, and use the Bonferroni correction for the family-wise error rate---dividing the typical significance threshold $\alpha=0.05$ by the number of regressions performed.

We perform a linear regression except for the case of instantaneous coordination, where we have reason to believe that a parabolic regression of the form $y=a(x+b)^2+c$, specifically with $a<0$ (an upside-down parabola) would be a better parameterization. We believe that while \textit{no} coordination is likely to produce unfavorable teaming, too-high values of \textit{instantaneous} coordination trade off with multi-turn coordination, which likely also produces unfavorable teaming, leading to a peak between the two.

In addition to the set of metric-by-metric regressions described in the main text, we also attempted analyses using a single multiple regression capturing all the candidate metrics. However, the strong correlations (multicollinearity) between metrics makes the fit unstable. Two other approaches were also considered, but ultimately rejected:

\begin{enumerate}
    \item \textbf{Feature selection based on variance inflation factor (VIF):} This approach attempts to downselect features to a subset that is less likely to cause multicollinearity issues \cite{lin2011vif}. However, in downselecting features, we lose much of our ability to analyze properties of individual features. For example, all game scores are highly correlated, but we showed that scores for inter-algorithm cross-play are much more spread out between agents, and only the highest-scoring agents exhibit a drop in teamwork rating.
    \item \textbf{Ridge regression:} Ridge regression is often used in cases of mutlicollinearity, as it penalizes the magnitude of the regression coefficient. However, using this method requires selecting an appropriate penalty, which is difficult to do in our setting. We found in unregularized least squares that the use of dominated moves tends to have a much higher coefficient (i.e. a small increase in the frequency has a strong effect on the teamwork rating) --- one to two orders of magnitude greater than other significant factors --- making appropriate selection of an appropriate penalty for ridge regression unclear. 
\end{enumerate}

\section{Extended Discussion on Objective-Subjective Metric Correlations}
\label{app:extended_obj_sub_correlation_analysis}

Since the AI-only data come from a different pool of games than the human-AI data, there is 1) much more AI-only data, and 2) not a direct mapping between data points in one pool to the other.
Thus, we chose to use the mean values of the AI-only metrics, and the distribution of values for the human-AI metrics, since the number of samples associated with the AI-only metrics can be made made arbitrarily large (due to inexpensive evaluation in comparison to human experiments). This choice results in heat maps that appear as ``strips'' associated with the mean AI-only metrics on the x-axis, but showing a distribution for the human response on the y-axis.

To show the potential pitfalls of the assumptions in the literature regarding correlations between game score and human-AI teaming, we first evaluate a correlation between game scores and teamwork rating, using 1) self-play games, 2) intra-algorithm cross-play games (games between agents using the same algorithm, but with different starting seeds; this excludes rule-based bots that do not use seeds), and 3) inter-algorithm cross-play games (games between agents using different algorithms).

Most of our information- and game-theoretic metrics show statistically significant correlations with teaming preference, but we note that our large sample size ($N=241$) contributed to low $p$ values and that the detection of the correlation ($p<0.05$) is not indicative of the \textit{size} of the correlation. For that, we further examined the correlation coefficient ($r$) and the slope of the regression ($m$).

\subsection{Task Performance}
\label{app:task_performance_metric_correllations}

\begin{figure}[h]
  \centering
  \includegraphics[width=0.9\linewidth]{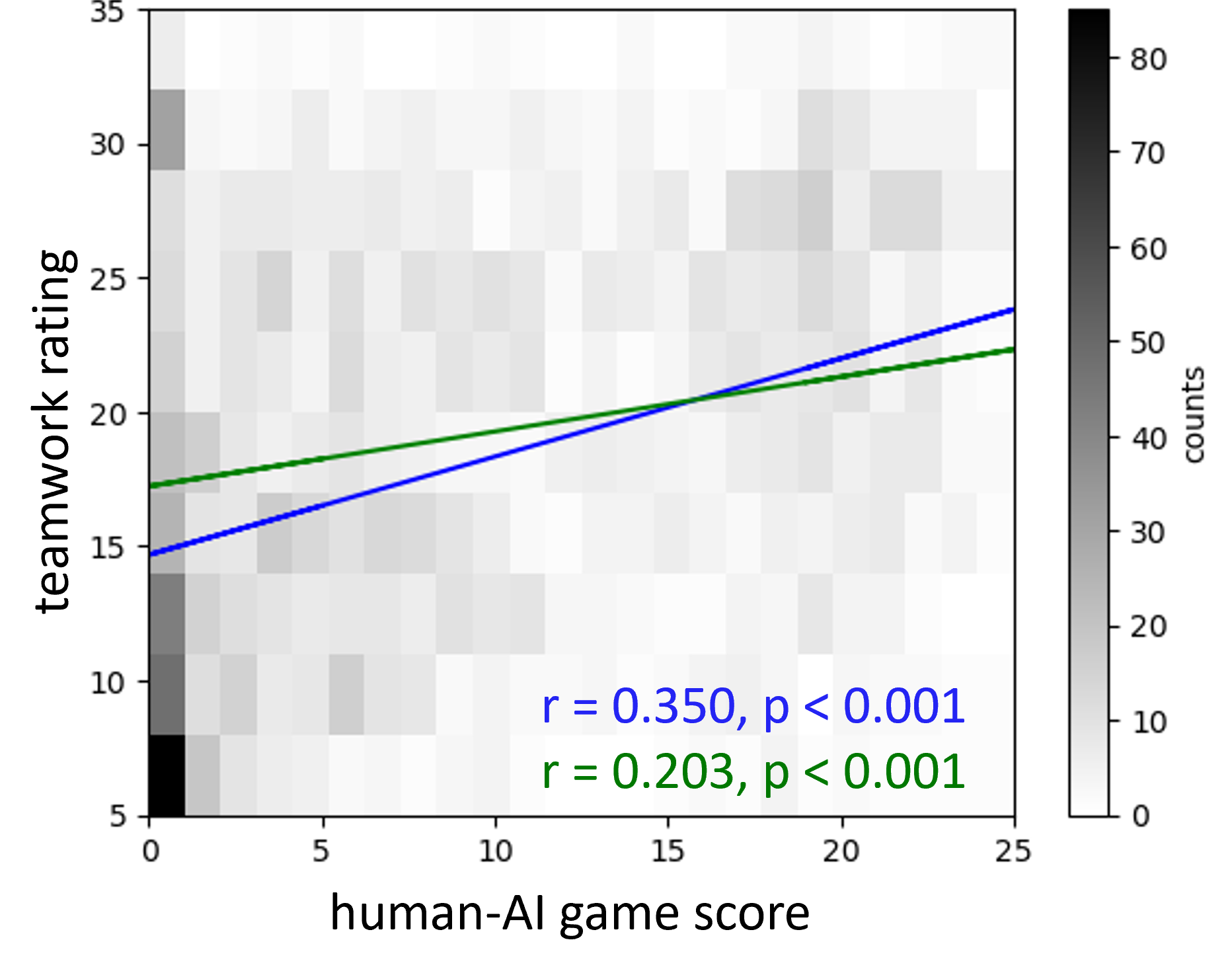}
  \caption{Scores during human-AI games vs human teamwork ratings. There is a medium to strong statistically significant correlation. Blue is data with RandomBot, green is without. The data from both the x and y axes come from the same games, so there is no need to take the mean of one side as in most other plots.}
  \label{fig:hai_game_score_vs_teamwork}
\end{figure}

First we test the hypothesis: \emph{humans prefer AI teammates for which they---as a human-AI team---achieve an overall higher game score} by correlating the teamwork rating with the human-AI game score as shown in Fig~\ref{fig:hai_game_score_vs_teamwork}. %
We see that the hypothesis is supported, with a statistically significant correlation between these variables, both with ($r=0.350$) and without ($r=0.203$) RandomBot included in the data. %

This result is notable as it provides empirical evidence on an open question raised in previous literature about the relationship between human preferences and teaming performance \cite{siu2021evaluation}. Our results indicate that humans tend to prefer AI teammates for which they achieve a higher game score, but the correlation is not as strong as one might expect; indeed Siu et al. \cite{siu2021evaluation} illustrates an exception to this finding where higher score did not correlate with a higher subjective rating of the AI. %
Of further note is the fact that human-AI game score---while being an \emph{objective} measure of teaming performance---is not measurable in the absence of human teammates. In contrast, AI-only objective metrics (Table~\ref{tab:agent_metrics}) that correlate to human preference could be much more valuable for AI-training purposes as they do not require the time-intensive process of incorporating a human in the loop.

\begin{figure*}[ht]
  \includegraphics[width=\linewidth]{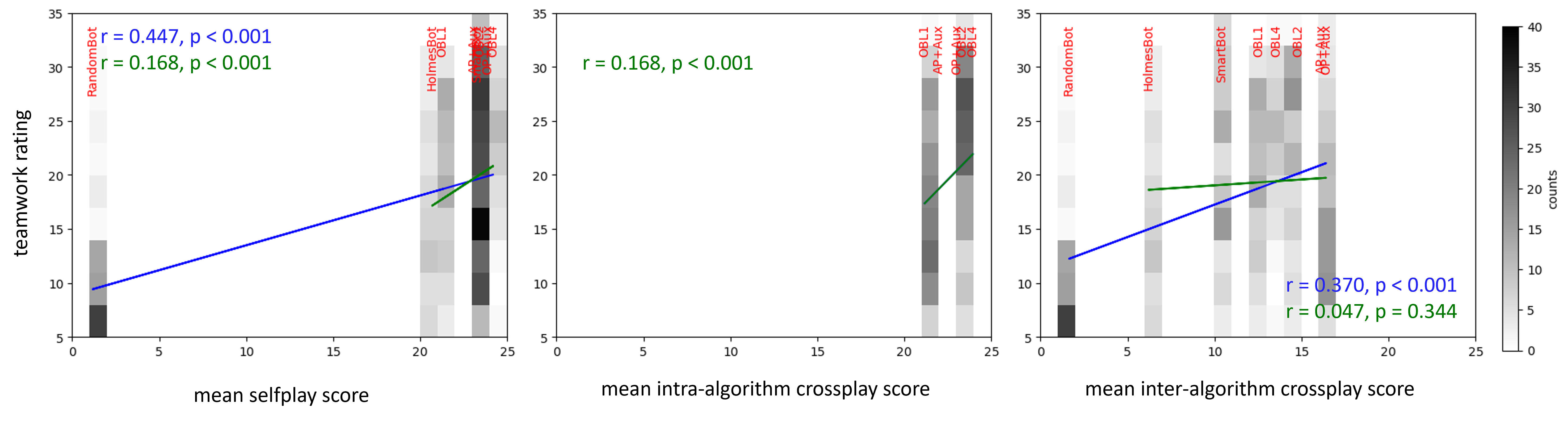}
  \caption{Teamwork rating vs self-play, intra-XP, and inter-XP scores. Correlation values match the colors of the associated lines (blue is with RandomBot, green is without). Intra-algorithm cross-play (intra-XP) cannot be evaluated on non-learning-based agents, thus rule-base bots (RandomBot, SmartBot, HolmesBot) do not appear in the second plot.}
  \label{fig:teamwork_vs_ai_game_scores}
\end{figure*}

To this end we next analyze the aggregate teamwork rating compared against AI-AI game scores; i.e. self-play, intra-XP, and inter-XP scores in Tables~\ref{tab:agent_metrics} and \ref{tab:ai_only_summary}. %
Hereby we are testing hypotheses that are implicit to much of the past work in Hanabi~\cite{lucas2022any,hu2020other} and reinforcement learning~\cite{strouse2021collaborating} research: i.e. \emph{humans prefer AI teammates that can achieve higher self-play, intra-XP, and inter-XP scores when evaluated in the absence of human teammates.}
Figure~\ref{fig:teamwork_vs_ai_game_scores} shows the teamwork rating plotted against different types of AI-AI game scores. Distributions of the Likert team ratings are plotted as heat map ``strips'' along the x-axis bin where the corresponding AI-only mean score is.

The correlation of score and teamwork rating is significant ($p<0.001$) in all cases of self-play (Figure~\ref{fig:teamwork_vs_ai_game_scores} left) and intra-XP (Figure~\ref{fig:teamwork_vs_ai_game_scores} middle), though the effect size $r$ is much less when excluding RandomBot. %
The slopes of the self-play lines indicate an increase in teamwork rating of $0.461$ and $1.049$ for every additional point in self-play (with and without RandomBot), and $1.648$ per additional point in intra-XP (Table \ref{tab:regressions}).

Ceiling and floor effects are an immediate concern about how to use these self-play and intra-XP scores: the scores are extremely tightly clustered and close to the maximum of 25 (or 0 in the case of RandomBot), limiting their predictive utility. Furthermore, a closer look at the distribution of teamwork ratings reveals that despite the positive correlation, the spread of ratings among agents with the scores (24 and 25) is quite large, with responses spanning nearly the entire Likert scale. More broadly, there are already concerns in the literature about using these same-agent (self-play) and similar-agent (intra-XP) evaluations, since they would allow for the formation of algorithm-specific conventions \cite{lucas2022any}. Finally, intra-XP score can only be evaluated in cases where we have agents that are trained on the same underlying algorithm, but with different training seeds, excluding from consideration any rule-based agents that do not use seeds.

Correlation of teamwork rating to inter-XP score (Figure~\ref{fig:teamwork_vs_ai_game_scores} right) is significant only when RandomBot is included, but RandomBot is also less of in outlier here because inter-XP scores are much lower and spread out than self-play and intra-XP scores.
The significant effect of the correlation disappears when we exclude RandomBot, and indeed becomes quite far from significant ($p=0.37$). In fact, the reason for the non-significant correlation can be seen: the agents with the \emph{highest} inter-XP scores (OP+Aux and AP+Aux) have markedly \textit{lower} teamwork ratings than the other agents.
It is worth noting that inter-XP score is highly dependent upon the AI agents present in the evaluation pool~\cite{lucas2022any}; Table~\ref{tab:bots_evaluated} represents the agent pool for our experiments. %
Thus it seems entirely possible to select an agent pool that would lead to more or less correlation between inter-XP and teamwork rating. %
However, to avoid cherry-picking data for desired results, a principled selection process for the agent pool would need to be developed. This process would require further justification and analysis of its own. %
Taken together, this all calls into question the utility of inter-XP as an evaluation or training metric for human-AI teaming agents and works to refute such assumptions in Lucas et al. \cite{lucas2022any}.

\subsection{Information-Theoretic}
\label{subapp:info_theory_metric_correllations}

\begin{figure*}[h]
  \centering
  \includegraphics[width=\linewidth]{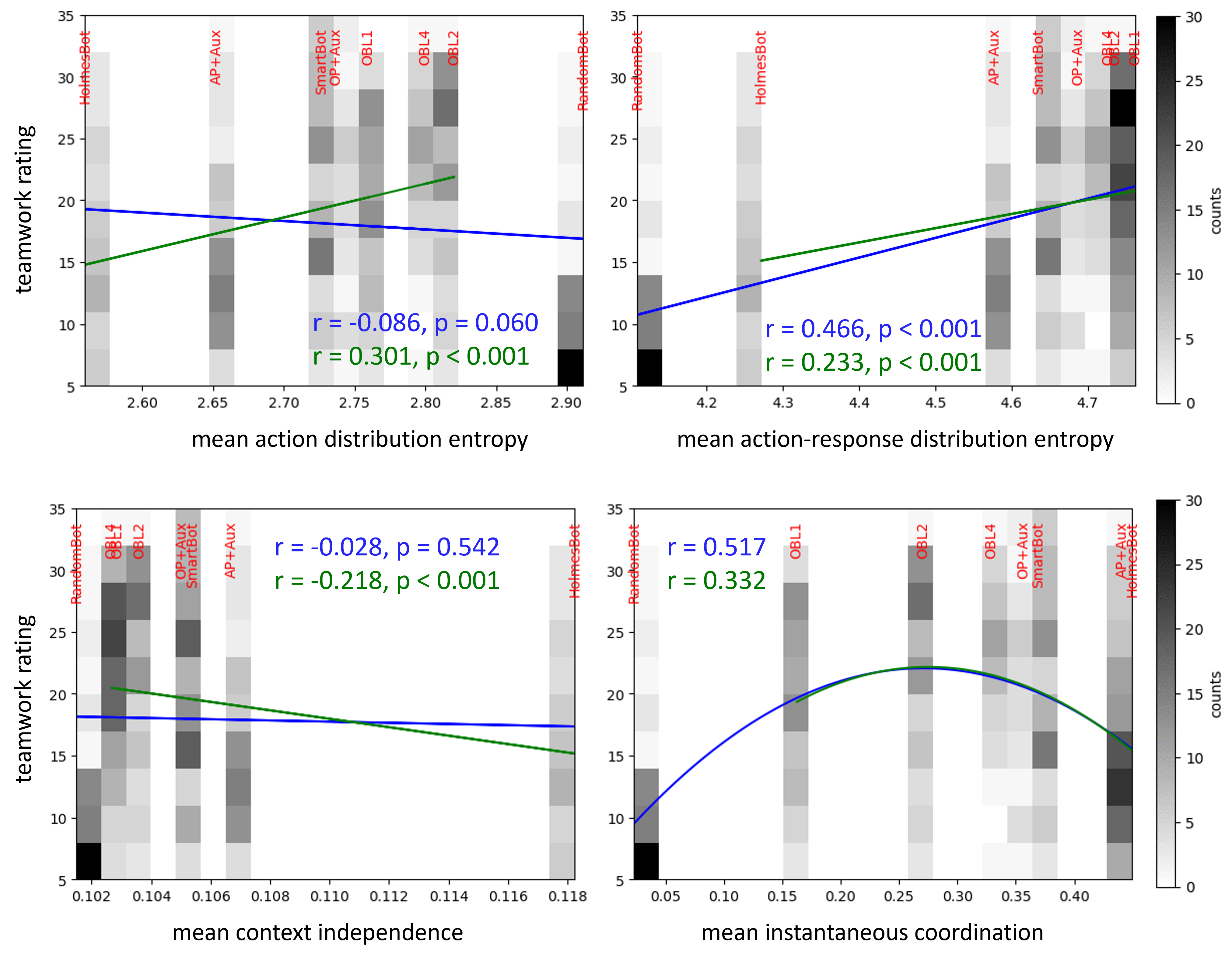}
  \caption{Information-theoretic metrics vs teamwork rating. Note that while the y-axes and colorbar are the same across all plots, the x-axes are not.}
  \label{fig:entropies_ic_ci}
\end{figure*}

Action distribution entropy 
(Fig~\ref{fig:entropies_ic_ci}, first) is a case where the inclusion or exclusion of RandomBot flips the correlation coefficient. Other than RandomBot, agents exhibit a moderate positive correlation ($r=0.301$).

Action-response distribution entropy 
(Fig~\ref{fig:entropies_ic_ci}, second) 
sees a very similar trend of agents as action distribution entropy, with almost the same ordering of agents along the entropy metric, but with the notable exception of RandomBot. The pool of agents that includes RandomBot shows a large correlation ($r=0.466$) and the pool without it exhibits a medium correlation ($r=0.322$), with both being statistically significant ($p<0.0001$). 

Context independence (Fig~\ref{fig:entropies_ic_ci}, third) exhibits a statistically significant negative correlation ($p < 0.001$, $r = -0.218$), when RandomBot is excluded.

Instantaneous coordination (Fig~\ref{fig:entropies_ic_ci}, last) is our only metric where we initially expect a nonlinear fit, and indeed, the data do show a parabolic pattern with a near-identical fit with and without RandomBot, showing strong ($r=0.515$) and moderate ($r=0.327$) correlation respectively. This regression is the strongest one (highest $|r|$) in our set for both cases.

For our entropy metrics, we should emphasize that despite the traditional association of entropy with randomness, perhaps a better interpretation in our context is action/action-response \textit{diversity}. RandomBot achieves high action diversity by acting uniformly randomly, but for \textit{action-response} entropy, RandomBot's extremely short games (Figure \ref{fig:turn_counts_AI_games}) concentrate its partner's responses to early-game moves, among which there is a much lower diversity of reasonable actions (e.g. typically giving hints), driving action-response entropy \textit{down}.

Apart from RandomBot's action entropy, the correlation of higher entropy (for both kinds) with greater human preference may indicate that more preferred agents use a greater variety of moves in general (action entropy) and in response to a partner's greater diversity of moves (action-response entropy). Greater entropy for both types may be indicative of a more sophisticated team strategy though these metrics are too coarse to probe strategy in any detail.

For instantaneous coordination, IC's peak shows the single-step coordination value that might be associated with a particularly good balance between short and long-term planning in Hanabi. Interestingly, here, AP+Aux and HolmesBot are approximately as far away from being human-preferred as RandomBot, showing the \textit{highest} level of instantaneous coordination. This feature may mean that these agents are heavily relying on short-term coordination, but in ways that humans are unable to understand and work with, a continuing problem in MARL settings that aim to eventually involve humans \cite{hu2020other,strouse2021collaborating}. While the data from this metric show some interesting features, it is unclear how it might be extended to other settings or used as a reward function in AI training, since the peak of human preference will likely change depending on the setting and on the actual strategy being employed.

The interpretation of context independence (CI) differs from its original presentation by Bogin et al. \cite{bogin2018emergence}. Given the way Hanabi is designed as a information-constrained game of \textit{implication} rather than explicit communication, \emph{lower} CI is hypothesized to correlate with better teamwork rating; this is in contrast to the original interpretation where higher CI indicates an agent's ability to leverage context as part of its communication. The CI value displayed by RandomBot is the lowest; which initially may be confusing because it would indicate that it is exhibiting the most context-\textit{dependent} behavior. This is the direction that we had argued would be occupied by the most \textit{preferred} agents---and indeed, we see that the OBL agents (a cluster of highly-preferred bots) is right next to RandomBot. But the location of RandomBot on the x axis is actually driven by the fact that its games are very short (it loses quickly), meaning the ``contexts'' in which it finds itself are very limited, lowering its CI value (Fig~\ref{fig:turn_counts_AI_games}). Apart from RandomBot, CI trends in the direction expected, with a negative correlation between the metric and teamwork rating; that is to say that, the more that an agents actions are dependent upon the state (context) of the game, then more human teammates prefere them. 
We note that---in spite of the statistically significant, medium correlation---the linear correlation model given here may not be the best fit for this data since we observe a rapid drop-off in teamwork rating with for low CI that appears to level-off with increased CI; perhaps representing an exponential decay model. 


\subsection{Game-Theoretic}
\label{subapp:game_theory_metric_correllations}

For analysis of game-theoretic objective metrics, we consider two types of actions by the AI agent. The first type represents \textit{short-term strictly dominated} actions---that is, actions that yield lower payoff in the current game turn than any other legal action, regardless of other players' actions. The second type represent \textit{short-term weakly dominant} actions---that is, actions that yield equal or higher payoff than any other legal action regardless of other players' actions \cite[Chp~4]{tadelis2013game}. %
These action types are defined over short time horizons so as to meet the bounded rationality criteria for human decision-making, and to make labeling them as such tractable \cite{simon1990bounded}.

\begin{figure*}[h]
  \includegraphics[width=\linewidth]{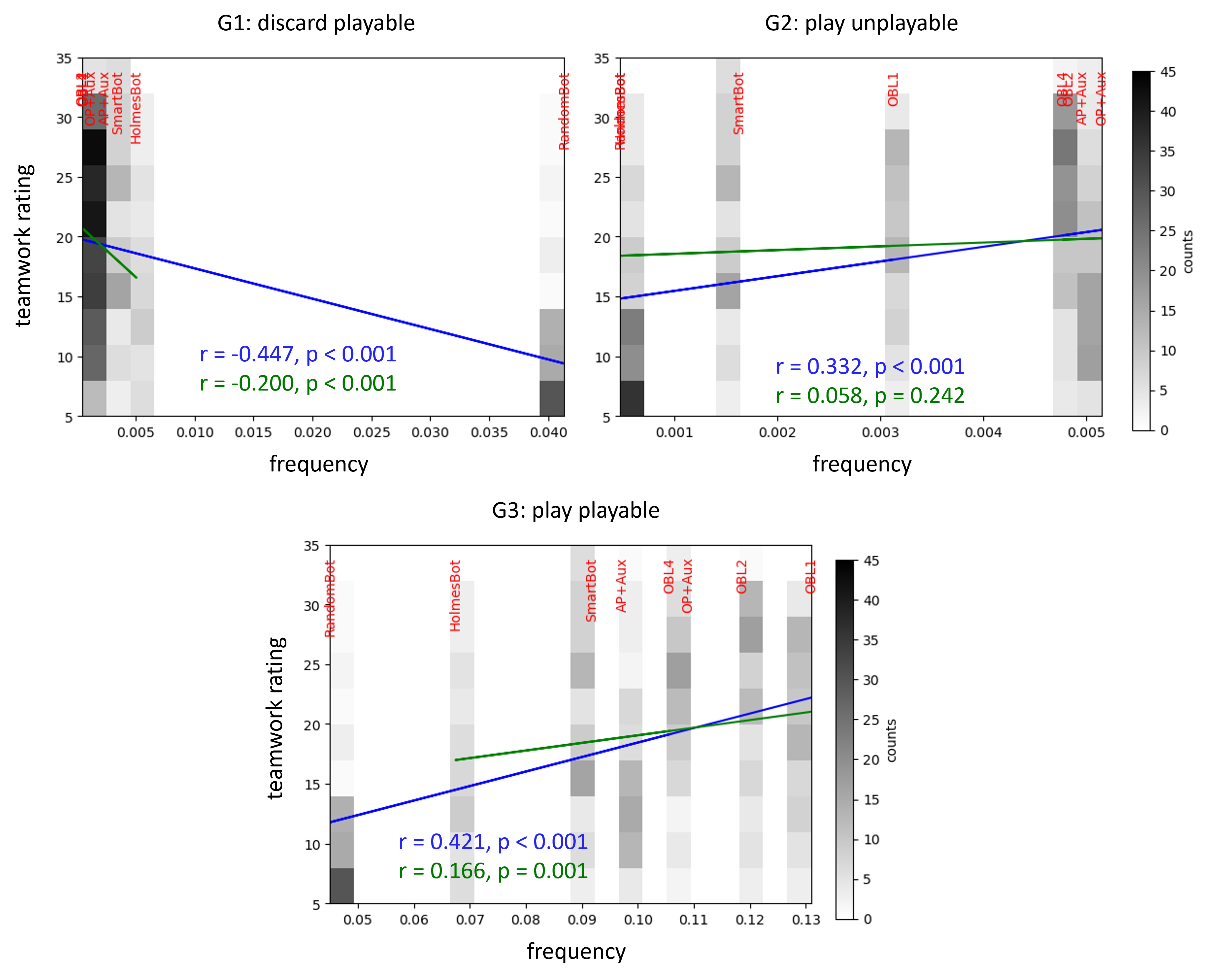}
  \caption{Frequency of discarding a card known to be presently playable (G1), playing a card known to be presently unplayable (G2), and playing a card known to be presently playable (G3) vs teamwork rating. G1 and G2 are domin\textit{ated} moves, while G3 is a domin\textit{ant} move. Note that while the y axes and colorbar are the same across all plots, the x axes are not.}
  \label{fig:G123_frequency}
\end{figure*}

Our first short-term strictly dominated move (G1, discarding a card that is known to be presently playable) shows a strong ($r=-0.447$) and moderate ($r=-0.200$) negative correlation with and without RandomBot, respectively (Figure~\ref{fig:G123_frequency} left). Notably, given that the x-axis is frequency, the slopes of the regressions ($m = -253$ with RandomBot, $m=-902$ without) are meaningful, and we see that even a slight increase in the frequency of this kind of action is correlated with a steep drop in human teamwork rating. In the case of the slope without RandomBot, we see that a 1\% increase in the occurrence of G1-dominated actions causes a 9-point drop in teamwork rating.

Our second short-term strictly dominated move (G2, playing a card that is known to be presently unplayable) only shows statistically significant correlation when RandomBot is included (Fig~\ref{fig:G123_frequency}, middle). Here, we note that both RandomBot and HolmesBot perform this kind of action most \textit{infrequently}. Likely, this is due less to their behavior in the face of card information (which RandomBot explicitly ignores), and more due to the short lengths of their games (Figure~\ref{fig:turn_counts_AI_games}). Very short games do not afford these bots enough information to 1) know that a card is presently unplayable, and 2) actually play it. In the absence of RandomBot, we see that the correlation is non-significant ($p=0.242$) and essentially flat ($r=0.058$), however, a separate calculation without \textit{both} RandomBot and HolmesBot gives a statistically significant ($p=0.003$) and negative ($r=-0.155$, $m=-651$) correlation, meaning that a 1\% increase in this kind of behavior (over longer games) results in approximately a 6.5 point drop in teamwork rating.



Both our dominated actions (G1 and G2) show negative correlations in most cases, and in all cases, give fairly large negative slopes, hinting at the large effects these actions have on perceptions of teamwork. It should be noted, however, that the use of these dominated actions suffers from the ``floor effect'', since non-random agents make such moves very rarely, so we come close to a minimum frequency of 0.0. G1 in particular has all agents other than RandomBot at extremely low values and clustered together.

Our weakly dominant action (G3, playing a card that is known to be presently playable) shows a significant positive correlation both with and without RandomBot (Fig~\ref{fig:G123_frequency}, right). Here, the slopes are $m=121$ with RandomBot, and $m=64$ without, indicating approximately a 0.64 to 1.21 point increase in teamwork rating for every percent increase in G3-dominant frequency. 

For our dominant action (G3-dominant), the absolute value of the slope is not nearly as large as the cases of the dominated actions, likely because it is only weakly dominant, and because dominat\emph{ed} actions represent negative events, which are more emotionally significant to players than the positive event of a domin\emph{ant} action due to negative psychological bias \cite{rozin2001negativity}. Notably, G3-dominant does not suffer from the same floor effect as the dominated actions, as the metric values are further from zero, and more spread out, giving us greater confidence in the utility of this metric and its correlation to teamwork.

\subsection{Post-Game Comparisons}
\label{app:post_game_comparisons}

\begin{figure}[h]
  \includegraphics[width=\linewidth]{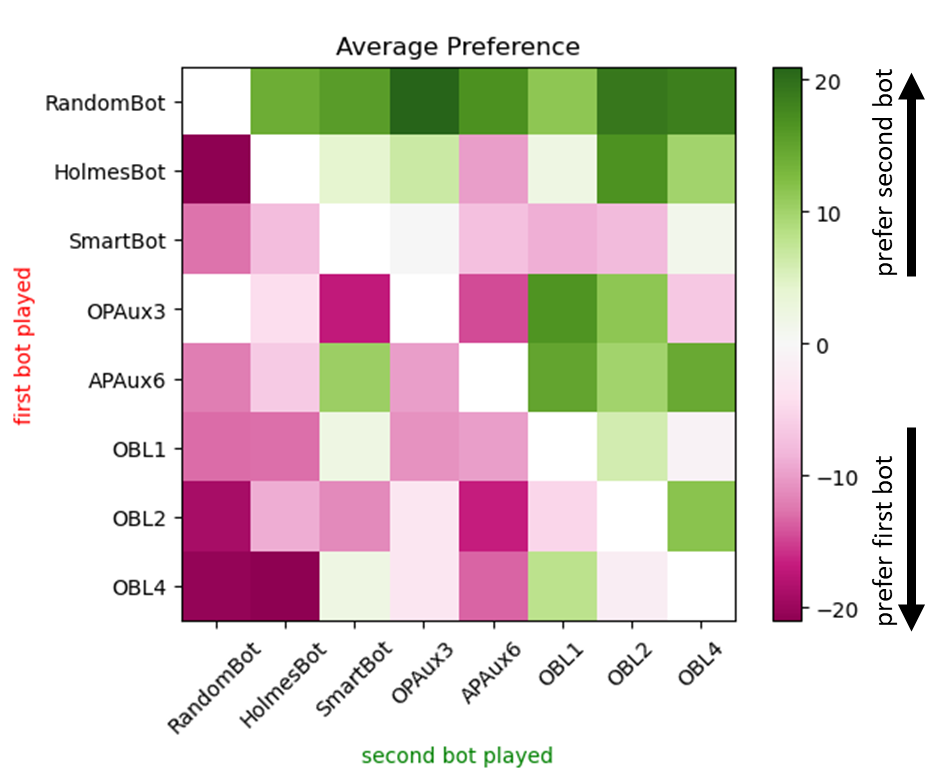}
  \caption{Pairwise subject preferences between the first and second bots they played with. Scores are a sum of Likert items P1 to P7, where the neutral value is taken to be zero for ease of visualization. Data do not exist on the diagonal since no bot was played with twice by the same person.}
  \label{fig:post_experiment_preferences}
\end{figure}

Subjects' post-game direct comparisons between bots is shown on Fig~\ref{fig:post_experiment_preferences}. Since the questions were asked in reference to the ``first bot'' and ``second bot'' the subjects played with, the results show a rough opposite symmetry along the diagonal --- bots had similar ratings regardless of whether they were experienced first or second.

\subsection{Other Data Products}

\begin{figure*}[h]
  \includegraphics[width=\linewidth]{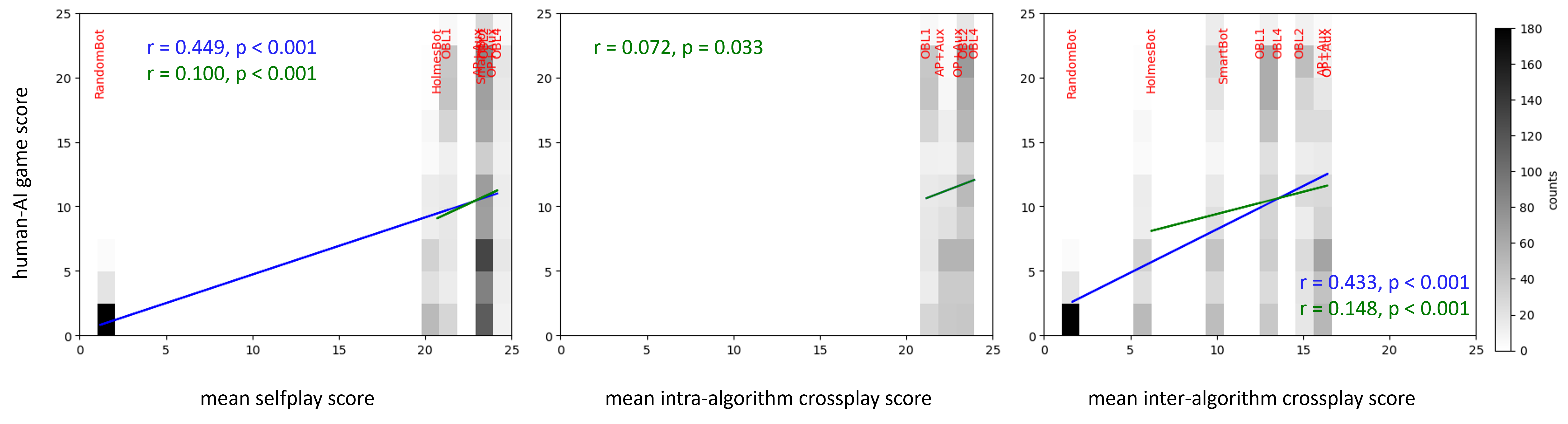}
  \caption{AI-AI scores vs human-AI scores. Though significant, we see that the correlations (without RandomBot as an outlier) are generally small, and notably, that the spread of human-AI game scores in the highest-performing AI agents is substantial. In fact, AP and OP agents show \textit{lower} human-AI game scores than the OBL agents, despite \textit{higher} AI-AI game scores.}
  \label{fig:scores_vs_hai_scores}
\end{figure*}

Though human-AI game scores are not the focus of our analysis, we show these correlations as well for the three types of AI-AI scores (Fig~\ref{fig:scores_vs_hai_scores}). These scores show a similar pattern to the human ratings of teamwork, though the inter-algorithm cross-play correlations here are significant (though still small). We see again that OP+Aux and AP+Aux agents show \textit{lower} scores than the OBL agents when teamed with humans, despite \textit{higher} scores in inter-algorithm cross-play.

\begin{figure}[ht]
    \centering
    \includegraphics[width=\linewidth]{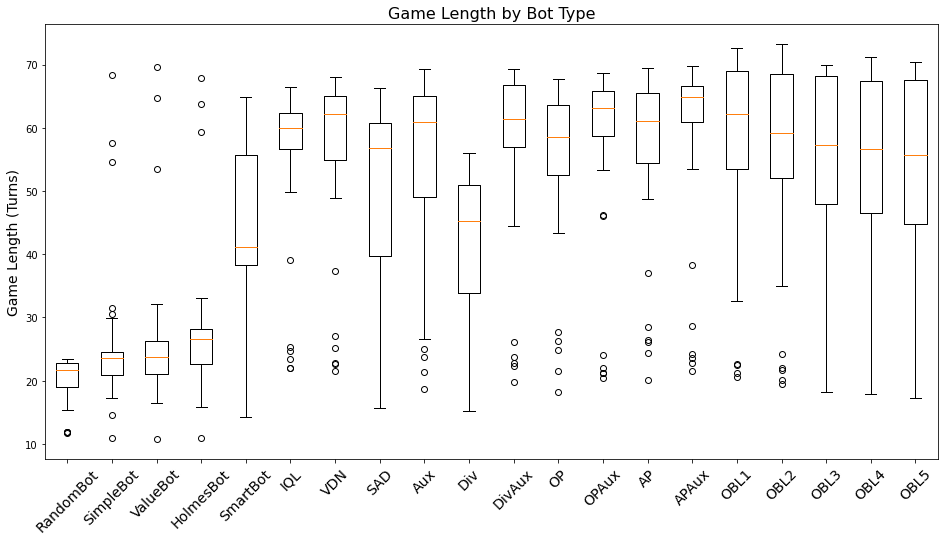}
    \caption{Number of turns in AI-only games.}
    \label{fig:turn_counts_AI_games}
\end{figure}

Figure~\ref{fig:turn_counts_AI_games} represents statistics for the number of turns within a single Hanabi game each AI agent tended to produce. This information is useful in understanding some of the counter-intuitive results for other objective metrics. For example, Section~\ref{subsec:sub_obj_metric_correlation} showed that RandomBot produced some of the lowest ARD-entropy and CI values, which is the opposite of what is expected for a completely random agent. Figure~\ref{fig:turn_counts_AI_games} helps explain this by showing that RandomBot also has the least number of turns per game, on average, because it ``bombs out'' so quickly, regardless of teammate. 

Other rule-based agents like SimpleBot, ValueBot, and HolmesBot have similarly short game length. SmartBot has the longest average game length for a rule-based bot. The learning-based bots, OP and OP+Aux, achieved considerably longer game lengths than SmartBot; this is of interest due to the direct comparisons made between SmartBot and OP+Aux in Siu et al. \cite{siu2021evaluation}. Some of the longest game lengths were achieved by AP+Aux; longer even than the human-preferred OBL bots.

\begin{figure}[ht]
  \includegraphics[width=\linewidth]{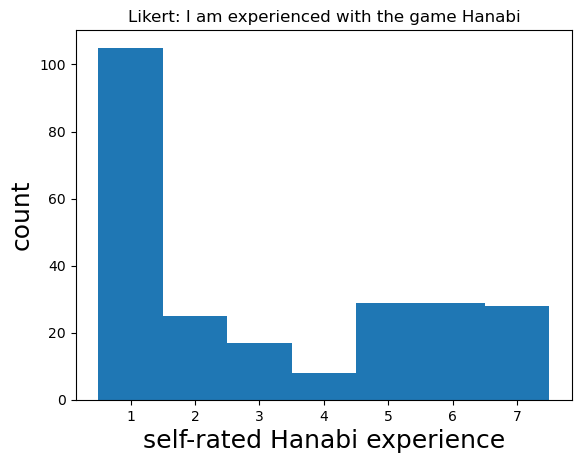}
  \caption{Subjects' self-rated familiarity with Hanabi (one of the demographic questions asked). There were a total of 241 subjects, of which 93 self-reported $\geq 4$ on familiarity.}
  \label{fig:familiarity}
\end{figure}

Figure~\ref{fig:familiarity} presents the self-reported participant familiarity with the game of Hanabi. This was important to assess because we wanted to ensure we were collecting a sufficient number of proficient Hanabi players. While there is a large number of completely inexperienced players (self-rated at 1 on Likert scale), there is almost an equal number of proficient-to-expert players (93 participants self-rated $\geq 4$)

By collecting this familiarity rating, it also allows for future data analysis that considers subjective teamwork rating broken down based upon the human participant's familiarity. For example, different AI characteristics (i.e. objective metrics) may correlate more strongly with teamwork rating when only considering ratings from expert Hanabi players than when considering ratings from novices.

\begin{figure}[h]
  \includegraphics[width=\linewidth]{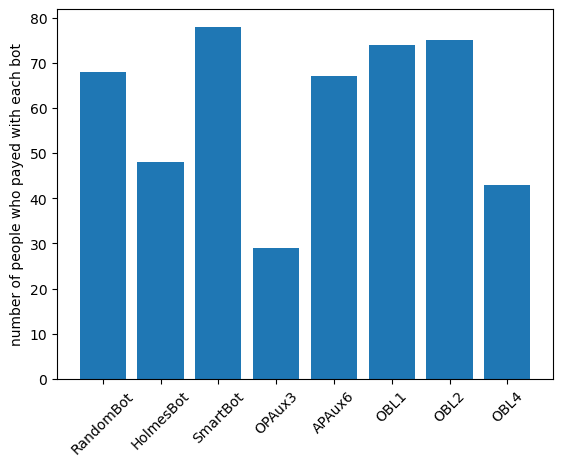}
  \caption{Distribution of participants per bot type. HolmesBot, OP, and OBL were added in a later round of experiments, resulting in fewer interactions.}
  \label{fig:participants_per_bot}
\end{figure}. 

\begin{figure}[h]
  \includegraphics[width=\linewidth]{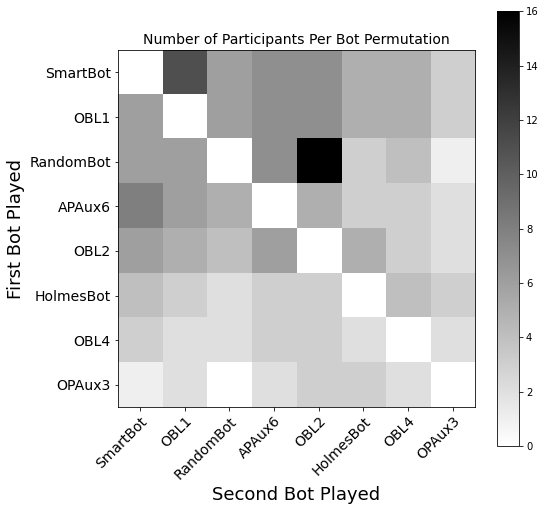}
  \caption{Distribution of participants per bot pairing.}
  \label{fig:bot_matrix}
\end{figure}

Figures~\ref{fig:participants_per_bot} and \ref{fig:bot_matrix} presents the number of human participants that interacted with each AI agent type. This information is tracked to ensure that we roughly balance between all agents to ensure the statistical power of our data is roughly equal between agents. We see that OP+Aux has the fewest interactions; this is due to it's late inclusion in the Human-AI teaming experiments

Figure~\ref{fig:bot_matrix} shows the frequency of pairs of AI agents played by a single human participant. For example, the SmartBot-row, OBL1-column shows a value of 16. This means that 16 human participants were assigned games just that SmartBot was the first AI teammate they encountered and OBL1 was the second. The diaganol is all zero because no participant was assigned the same AI for both teammates. The upper left corner is more densily filled because HolmesBot, OBL4, and OP+Aux were all introduced to the Human-AI experiments late in the process.

\end{document}